\documentclass[12pt, a4paper]{article}
\pdfoutput=1
\usepackage{amsmath}
\usepackage{amsfonts}
\usepackage{amssymb}
\usepackage{graphicx, rotating}
\usepackage{epsfig}
\usepackage{latexsym}
\usepackage{graphicx}
\usepackage{color}
\usepackage{amsmath,bm,amssymb}
\usepackage{cite}
\usepackage{slashed}
\usepackage{hyperref}
\hypersetup{colorlinks, citecolor=bluscuro, linkcolor=black, urlcolor=bluscuro}
\definecolor{rossos}{cmyk}{0,1,1,0.55}
\definecolor{bluscuro}{rgb}{0.15, 0.2, .85}
\definecolor{bluchiaro}{cmyk}{1,.3,0.,0.1}

\def\simlt{\stackrel{<}{{}_\sim}}

\numberwithin{equation}{section}

\newcommand{\nn}{\nonumber}

\newcommand{\be}{\begin{equation}}
\newcommand{\ee}{\end{equation}}
\newcommand{\bea}{\begin{eqnarray}}
\newcommand{\eea}{\end{eqnarray}}

\newcommand{\arXiv}[2]{\href{http://arxiv.org/pdf/#1}{{\tt [#2/#1]}}}
\newcommand{\arXivold}[1]{\href{http://arxiv.org/pdf/#1}{{\tt [#1]}}}

\def\bma#1{\mbox{\boldmath{$#1$}}}

\begin{document}
\allowdisplaybreaks
%FRONTPAGE2%%%%%%
\begin{titlepage}
\begin{flushright}
%CERN-TH-2023-XYZ
\end{flushright}
\vspace{.3in}

\vspace{1cm}
\begin{center}
{\Large\bf\color{black} 
An Exploration of Vacuum-Decay Valleys} \\
\vspace{1cm}{
{\large J.R.~Espinosa$^{a}$ and T. Konstandin$^{b}$} 
\vspace{0.3cm}
} \\[7mm]
{\it {$^a$\,  Instituto de F\'{\i}sica Te\'orica, IFT-UAM/CSIC, \\ 
C/ Nicol\'as Cabrera 13-15, Campus de Cantoblanco, 28049, Madrid, Spain}}\\
{\it $^b$ {Deutsches Elektronen-Synchroton DESY, Notkestr. 85, 22607 Hamburg, Germany}}
\end{center}
\bigskip

\vspace{.4cm}

\begin{abstract}
In the standard lore the decay of the false vacuum of a single-field potential is described by a semi-classical Euclidean bounce configuration that can be found using overshoot/undershoot algorithms, and whose action suppresses exponentially the decay rate. While this is generically correct, we show in a few concrete examples of potentials, previously studied in the literature for other purposes, that the vacuum decay structure can be far richer. In some cases there is no bounce and decay proceeds via the so-called pseudo-bounce configurations. In the general case with bounce, there are $2n+1$ bounces,
with $n$ ranging from 0 (the standard case) to $\infty$.  Some of these decay configurations we call antibounces as they have the wrong behavior for overshoot/undershoot algorithms, which can miss them. Bounce and antibounce configurations form $n$ pairs  connected by pseudo-bounces. Our analysis benefits from a combined use  of Euclidean and tunneling potential methods.
\end{abstract}
\bigskip

\end{titlepage}

\section{Introduction}

The decay of metastable states is ubiquitous in cosmology, string theory and particle physics.
Usually, the semi-classical bounce configuration dominates the tunneling amplitude,  
which is exponentially suppressed with the action of this configuration~\cite{Coleman}. The bounce can be found using a standard overshoot/undershoot algorithm. 

In this paper we show that, for some potentials, the situation can be considerably richer. There are potentials for which there is no bounce to describe vacuum decay (as has been already studied in the literature), while for others more than one bounce (even an infinite number of them) can exist.\footnote{See \cite{HW,Masoumi,Arttu,BoN2} for an incomplete list of previous work discussing the possibility of having more than one bounce (for single-field potentials beyond the special scale invariant $V=-\lambda\phi^4$ which has an infinite number of degenerate bounces). Note however that all of these works except \cite{Masoumi} include gravitational effects important for the bounce structure, while we work with gravity decoupled.} In this study we use a combination of Euclidean techniques plus the alternative tunneling potential method (reviewed in section~\ref{sec:Vt}). We also rely on the analysis of pseudo-bounce configurations, which were introduced in \cite{PS} precisely to describe decays when there is no bounce, although they are of more general interest (we give more details on these configurations in section~\ref{sect:ps}).

A pseudo-bounce Euclidean configuration, $\phi(r)$, has a constant inner core [$\phi(r) = \phi_e$ = const for $r<r_i$] 
and then follows the bounce equation of motion. Tuning the core size, this solution will fulfill
the boundary conditions for $r \to \infty$. That this indeed describes a physical tunneling amplitude,
is seen by modifying the potential: The pseudo-bounce describes a conventional bounce in 
a potential that is lifted beyond $\phi_e$, in such a way that there is a cusp and  a local minimum of the potential at $\phi_e$. This ensures many desirable properties of the pseudo-bounce,
for example that the nucleated bubbles from the pseudo-bounce conserve energy.

Pseudo-bounces can also be described using the 
tunneling potential method~\cite{E}. One nice feature of this method is that 
the core of the pseudo-bounce does not have to be treated separately. 
The action of the core is already baked into the usual expression for the 
action of the tunneling potential method.
Moreover, pseudo-bounces can be easily found in the tunneling potential 
method by just modifying the initial conditions~\cite{PS}.

From the point of view of the tunneling potential method, pseudo-bounces are the answer to the question of
which configuration minimizes the action if we hold fixed the value at its center (while a bounce minimizes
the action without that restriction and thus gives the global minimum of the action). This point of view
highlights the fact that pseudo-bounces are local minima of the action functional in slices of configuration space with fixed $\phi_e$. In other words, pseudo-bounces trace the bottom of valleys of the action functional in configuration space. This paper can then be understood as an exploration of such decay valleys. This exploration will be done on some concrete examples: single-field potentials that were already discussed in previous literature and  happen to have a very rich structure of decay channels. The use of pseudo-bounces is crucial to shed light on such more general decay structures.

In section~\ref{sect:exD} we analyze a potential introduced for other purposes in \cite{EK}, showing that it features 5 bounces, 2 of which we classify as "antibounces" as they do not behave in the standard way when faced with an overshoot/undershoot algorithm. Indeed, raising their central value $\phi(0)$ leads to an undershot
while lowering it leads to an overshot (the opposite of what one expects of a standard Coleman bounce).

In section~\ref{Sasaki} we analyze a potential introduced in \cite{Sasaki} that features a singular bounce, with $\phi(0)=\infty$, but finite action. We discover that, besides that singular bounce, the potential admits an infinite number of bounces (and antibounces).

In the two previous examples the potentials are unbounded from below and there are decay modes with arbitrarily low action, so that the false vacua are badly unstable. To show that unboundedness is not required to get the rich structure of  decay channels these models display, in section~\ref{sect:exDreg} we modify the potential of section~\ref{sect:exD} regularizing it so that it has a finite global minimum at a finite field value. The main properties of the rich decay structure of the unregularized potential remain, but the decay action is now finite.

Section~\ref{sect:genless} collects some general lessons we draw from the study of our examples, while Appendix~A gives details of the singular bounce of the potential of section~\ref{Sasaki} and Appendix~B discusses the possible implications of crossing tunneling potentials.

\section{Review of the Tunneling Potential Approach\label{sec:Vt}}

In this section we briefly summarize the main features of the tunneling potential formalism, proposed in \cite{E} to describe semiclassical false vacuum decay in an alternative way that does not involve the Euclidean quantities of Coleman's approach \cite{Coleman}. For simplicity we focus on $4d$ single-field models and do not include gravitational corrections, although the formalism can be extended to both cases, see \cite{Eg,EKmulti}.

Consider the decay of the false vacuum at $\phi_+$ of some potential $V(\phi)$. The calculation of the tunneling action for such decay in the tunneling potential approach takes the following (variational) form: find the tunneling potential function $V_t(\phi)$, that connects $\phi_+$ to some $\phi_0$ on the basin of the true vacuum\footnote{We assume $\phi_->\phi_+$, so that $\phi_+<\phi_0< \phi_-$. Usually we set $\phi_+=0$ and $V(\phi_+)=0$.} at $\phi_-$, and minimizes the action functional \cite{E}
\be
 S[V_t]=54\pi^2\int_{\phi_+}^{\phi_0} \frac{(V-V_t)^2}{-{V_t'}^3}d\phi\ ,
\label{SVt}
\ee
where a prime denotes a field derivative. The minimal value of $S[V_t]$ coincides with the tunneling action found in the Euclidean formalism \cite{Coleman} for the bounce solution, an extremal (a saddle point) of the Euclidean action. The $V_t$ approach has many good properties
that have been discussed elsewhere (see {\it e.g.} \cite{Unreasonable} and references therein). 

The Euler-Lagrange equation $\delta S[V_t]/\delta V_t=0$ reads
\be
(4V_t'-3V')V_t' + 6(V-V_t)V_t''=0
\ ,
\label{EoMVt}
\ee
and gives us the "equation of motion" (EoM) for $V_t$.
The solution $V_t$ "tunnels" under the potential barrier ($V_t\leq V$), it is monotonically decreasing from $\phi_+$ to $\phi_0$, and has boundary conditions
\be
V_t(\phi_+) = V(\phi_+), \quad V_t(\phi_0) = V(\phi_0), \quad
V_t'(\phi_+)=V'(\phi_+)=0,\quad V_t'(\phi_0)=\frac34 V'(\phi_0)\ .
\label{MinkBCs}
\ee
The field value $\phi_0$ must be determined so as to satisfy the previous boundary conditions and coincides with the value of the Euclidean bounce at its center. Thus, the field interval involved in the tunneling process is the same in both formalisms, as one would expect.

A dictionary between the Euclidean and tunneling potential formalisms to translate results between the two is useful. In Coleman's Euclidean approach, false vacuum decay is described by the so-called Euclidean bounce, an $O(4)$-symmetric configuration $\phi(r)$, that extremizes the Euclidean action. The bounce is thus
a solution of the Euler-Lagrange equation
\be
\ddot \phi +\frac{3}{r}\dot\phi =V'\ ,
\label{EoMphi}
\ee 
which can be interpreted as describing the motion of a point particle with "position" $\phi$ at time "r" sliding down an inverted potential $-V$ and subject to a "velocity"-dependent friction force, with boundary conditions $\phi(0)=\phi_0$, $\dot\phi(0)=0$, $\phi(\infty)=\phi_+$.

The key relation between both formalisms is
\be
V_t (\phi)= V(\phi) -\frac12 \dot\phi^2\ ,
\label{Vtlink}
\ee
where $\dot x\equiv dx/dr$, and $\dot\phi$ is expressed in terms of the field using the bounce profile $\phi(r)$. 
From~(\ref{EoMphi})
one further gets the relations
\be
\dot\phi = - \sqrt{2(V-V_t)}\ ,\quad \ddot\phi=V'-V_t'\ ,
\label{dphi}
\ee
where the minus sign for $\dot\phi$ follows from our convention $\phi_+<\phi_-$. 

On the other hand, knowing $V_t$, the bounce field profile can be obtained from the previous formulas. 
Another useful relation gives the Euclidean radial distance from the center of the bounce in terms of $V$ and $V_t$ as\cite{E}
\be
r=\frac{3\sqrt{2(V-V_t)}}{-V_t'}\ ,
\label{rVt}
\ee
which follows directly from the Euclidean EoM for the bounce (\ref{EoMphi}) and previous relations.

\section{Pseudo-Bounces\label{sect:ps}}

It is easy to construct potentials with false vacua that cannot decay via Coleman bounces, cases in which the bounce equation has no solution satisfying the correct boundary conditions. 
Perhaps the simplest example is a negative quartic potential perturbed by a mass term, $V(\phi)=-\lambda \phi^4/4+m^2\phi^2/2$. With $m=0$, scale invariance causes the action to have a flat direction in field configuration space, with action $S=8\pi^2/\lambda$ and arbitrary $\phi_0$ (so that there are infinite bounces with degenerate action). The field value $\phi_0$ can be taken as a coordinate along that flat direction. For $m^2>0$ however, all trial solutions for the bounce EoM are undershots and the previous flat direction is lifted into a runaway direction in field configuration space with the bounce ``pushed to infinity'' \cite{PS}.\footnote{For $m^2<0$ all trials are instead overshots and the decay action can be made arbitrarily small \cite{PS}.}

Such vacua decay nevertheless via different field configurations [still $O(4)$ symmetric] called pseudo-bounces in \cite{PS}. In the Euclidean formalism of Coleman, pseudo-bounces have a homogeneous inner core with radius $r_i$ where the field takes a constant field value $\phi_e$.\footnote{In Euclidean language, the value $r_i$ can be understood as a "waiting time"
for friction to get reduced, thanks to which an undershot solution can reach the false vacuum.} Outside the core, the field tends toward the false minimum of the potential as a solution of the bounce equation. Although these configurations are not proper bounces (thus the name) they enjoy some of the good properties of  bounces. In particular, 1) the slice of the pseudo-bounce 
at zero Euclidean time that gives the nucleated three-dimensional bubble mediating decay has zero energy, so that energy is conserved in the decay; and 2) they give the lowest value of the tunneling action for fixed $\phi_e$. We refine this naive expectation in this paper. Due to this last property, one can think of pseudo-bounces as configurations that track the bottom of a sloping-valley in field configuration space (as in the example mentioned above), with $\phi_e$ as coordinate and with the true bounce only reached as $\phi_e\to\infty$. 

Interestingly, pseudo-bounces also exist when the potential does have a proper bounce (say with central value $\phi_0$). In the standard case, pseudo-bounces appear for $\phi_e<\phi_0$, where one
normally has undershots, with their action monotonically decreasing towards the minimum at the proper bounce, which can be found using the standard overshoot/undershoot method.  In such case, pseudo-bounces also track the bottom of a valley, but the true bounce is reached at some finite $\phi_e$.
Because pseudo-bounces have actions larger than the proper bounce they are subleading for decay, which is dominated by the true bounce. Nevertheless, they can still play a role if the slope of the valley approaching the true bounce is small, so that they are not much suppressed compared to the bounce.

The tunneling potential method finds naturally such pseudo-bounce solutions by solving (\ref{EoMVt}) with the boundary condition $V_t'(\phi_e)=0$, instead of the bounce condition $V_t'(\phi_0)=3V'(\phi_0)/4$. This is directly connected with the fact that the radius of the inner core of the Euclidean pseudo-bounce profile is given by \cite{PS}
\be
r_i=\lim_{\phi\to\phi_e}\frac{3\sqrt{2(V-V_t)}}{(-V_t')}\neq 0\ .
\ee
We show in later sections how the $V_t$ formalism is ideally suited to find whole families of pseudo-bounce configurations. 

Before closing this section, let us remind the reader of
one of the nice features of the formalism to deal with pseudo-bounces: it takes into account automatically the core contribution to the action of the pseudo-bounce ~\cite{PS}. 
In the Euclidean description the tunneling action for a pseudo-bounce gets three contributions: the core term and the gradient and potential terms from outside:
\be
S_E = S_C + S_K + S_P \, ,
\ee
with
\bea
S_C &=& 2\pi^2 \int_0^{r_i} dr \, r^3  \, V(\phi_e) = \frac{\pi^2}{2} r_i^4 V(\phi_e) \, ,\nn\\
S_K &=& 2\pi^2 \int_{r_i}^\infty dr \, r^3  \, \frac12 \dot\phi^2  \, ,\nn \\
S_P &=& 2\pi^2 \int_{r_i}^\infty dr \, r^3  \, V(\phi)  \, . 
\eea
Integrating the potential term by parts and using the equation of motion (\ref{EoMphi})
yields $S_P = - S_C - S_K/2$, such that (on-shell) one finds $S_E = S_K/2$.
The action for the potential method (\ref{SVt}) translates directly into this gradient term and hence
takes the core term already properly into account.

%%%%%%%%%%%%%%%%%%%%%%%%%%%%%%%%%%%%%%%%
\section{Example from \cite{EK}\label{sect:exD}}
We take our first example potential with a rich decay structure from \cite{EK}.
In that paper, several one-field potentials admitting an analytic solution for vacuum decay were presented.
We examine example D of \cite{EK},
\be
V(\phi)=\mathrm{Ei}(\log\phi^2)+\frac16 \phi^2\left(1-\frac{\log^2\phi_0^2}{\log^2\phi^2}\right)\ ,
\label{VexD}
\ee
where $0<\phi_0<1$ and $\mathrm{Ei}(x)$  is the exponential integral function. This potential has a false vacuum at $\phi_+=0$.
An analytic Euclidean bounce describing the decay of the $\phi_+$ vacuum was calculated exactly in \cite{EK} as
\be
\phi_B(r)=e^{-\sqrt{r^2/3+L_0^2}}\ ,
\label{phiB09}
\ee
where $L_0\equiv\log\phi_0$, so that $\phi_0$ corresponds to the central value of the bounce, $\phi_B(0)$. The tunneling potential associated to this bounce is
\be
V_t(\phi)=\mathrm{Ei}(\log\phi^2)\ ,
\label{Vt09}
\ee
and the tunneling action can be obtained analytically as \cite{EK}
\be
S=\frac{3\pi^2}{16}\left[\phi_0^2(3-6L_0+2L_0^2+4L_0^3)-8L_0^4\mathrm{Ei}(2L_0)\right]\ .
\label{San}
\ee

\begin{figure}[t!]
\begin{center}
\includegraphics[width=0.6\textwidth]{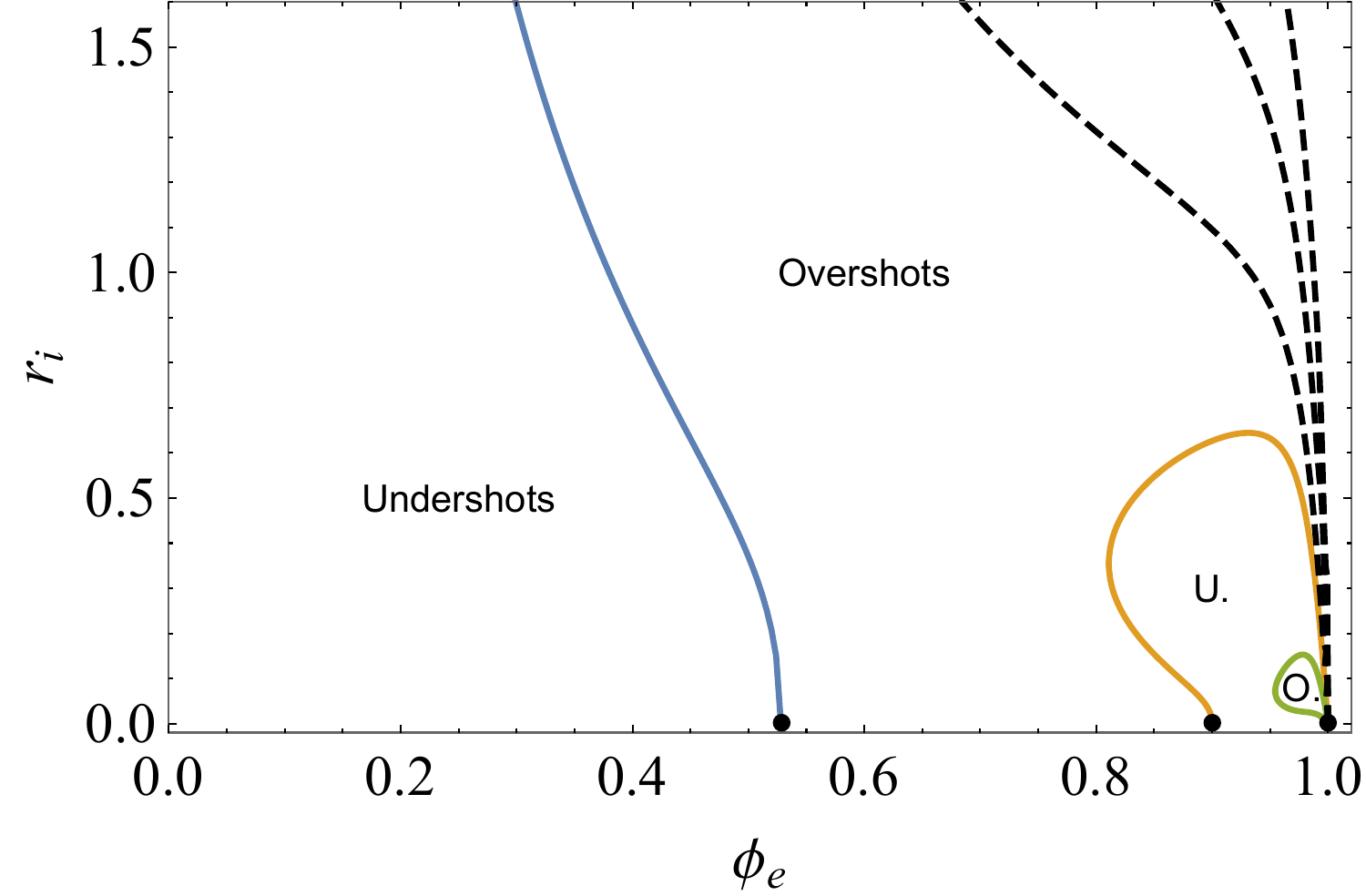}
\end{center}
\caption{For the example of section~\ref{sect:exD}, with $\phi_0=0.9$, different pseudo-bounce branches (with  core field value $\phi_e$ and  inner core radius $r_i$) separating overshot and undershot regions. Black-dashed lines are multi-pass pseudo-bounces. True bounces live on the $r_i=0$ axis and are indicated by black dots. 
\label{fig:ExDri}
}
\end{figure}

\subsection{(Euclidean) Pseudo-bounce Solutions}
For the numerical analysis, consider the case with $\phi_0=0.9$. To get a pseudo-bounce solution with a given 
central value $\phi_e$ in the Euclidean approach,  the core radius $r_i$ has to be tuned to a particular value between overshots and undershots. Figure~\ref{fig:ExDri} shows the distribution of overshots and undershots in the plane 
$(\phi_e,r_i)$. The lines separating undershots from overshots correspond to pseudo-bounce solutions, with a gap in the possible values of $\phi_e$, for $\phi_e\in (0.53,0.81)$,
where only overshots are obtained. On the other hand, for some ranges of $\phi_e$ more than one pseudo-bounce solution exists. 

Black-dashed lines correspond to pseudo-bounces that
overshoot but come back afterwards to the false vacuum. We call these solutions multi-pass pseudo-bounces.\footnote{These solutions exists e.g. if the potential is symmetric, $V(\phi)=V(-\phi)$,
with the false vacuum at $\phi=0$, so that there are in fact three vacua, at $\phi=\{0,\pm\phi_-\}$. They do not exist if the potential has just two vacua.\label{Vsym}} The lowest one in the figure crosses the false vacuum only once, the second lowest
twice and so on (we only show the lowest three in the figure). These solutions typically cost higher Euclidean action and one expects them to feature more than a single negative mode and are thus of rather limited interest.

As $r_i\to 0$, the family of pseudo-bounce solutions approaches a Coleman bounce solution (marked by a black dot). We see that there are several Coleman bounces: 
the one with $\phi_e=\phi_0=0.9$ is the analytical one studied in \cite{EK} and given in (\ref{phiB09}), but there are  more: one with $\phi_e\simeq 0.53$ and three (leaving aside multi-pass ones) with the same asymptotic boundary condition $\phi_e= 1$. 
Let us examine these bounces in turn.

\begin{enumerate}
\item The standard overshoot/undershoot search algorithms would easily find the bounce with $\phi_e\simeq 0.53$ as it has the usual behaviour: larger (smaller) $\phi_e$ leads to an overshot (undershot), in agreement with the behaviour described in Coleman's seminal paper \cite{Coleman}. The pseudo-bounce line attached to this bounce goes to $r_i\to\infty$ at some finite field value $\phi_\infty$, corresponding to $V(\phi_\infty)=V(\phi_+)$ (only with zero friction $\phi_\infty$ could be connected to $\phi_+$). This configuration has infinite radius and infinite action.

\item The analytic bounce at $\phi_e=0.9$, however, behaves in the opposite way: larger (smaller) $\phi_e$ leads to an undershot (overshot) and could easily escape a naive search. 
In the rest of the paper we refer occasionally to such bounces as antibounces, for lack of a better name.
It might seem at first glance that such behaviour should not be possible: if the solution starts at a $\phi_e$ higher than that of the bounce ($\phi_{Be}=0.9$), some $r$-time is needed to reach $\phi_{Be}$, larger $r$ reduces the friction and naively one would conclude that one is thus forced to have an overshot. However, the friction term in the Euclidean equation is proportional to $\dot\phi/r$: it decreases with larger $r$ but increases with larger $\dot\phi$. Then, having a large potential slope (leading to larger $\dot\phi$) can compensate for the larger $r$. That is why we can have undershots
for $\phi_e>\phi_{Be}$ and large potential slopes are needed for this to happen. We discuss this point further below.

\item Finally, the three bounces at $\phi_e=1$ (one for each pseudo-bounce line reaching $\phi_e=1$) are hard to find because the potential is singular at $\phi_e=1$. However, this peculiarity is less relevant: a singularity of the potential at a finite field value is not usually acceptable. Nevertheless, the presence of an antibounce with the opposite undershot/overshot behaviour (as in point 2 above) can appear in normal potentials without singularities.
\end{enumerate}

\subsection{Tunneling Action $\bma{S(\phi_e)}$}
\begin{figure}[t!]
\begin{center}
\includegraphics[width=0.45\textwidth]{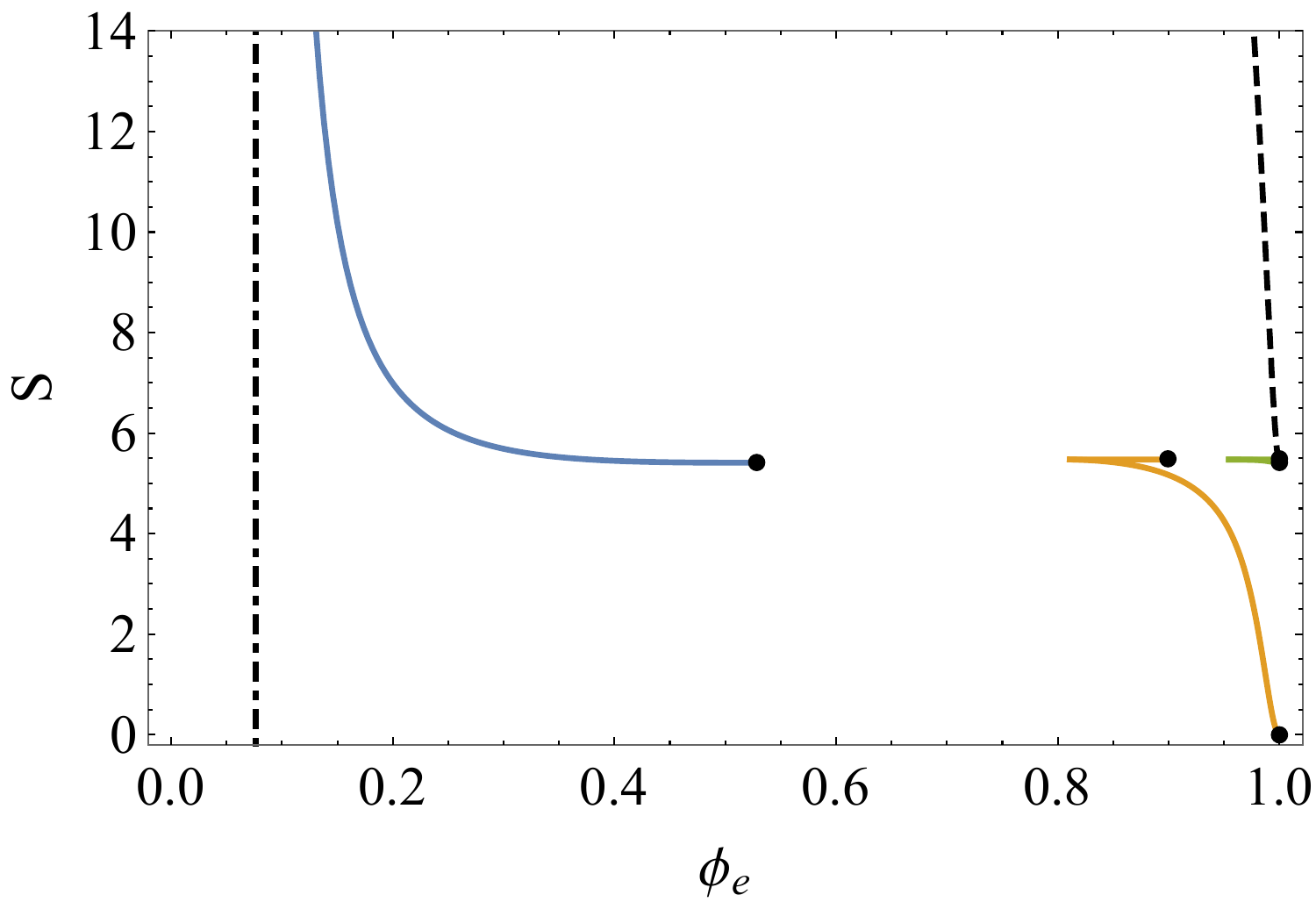}
\includegraphics[width=0.45\textwidth]{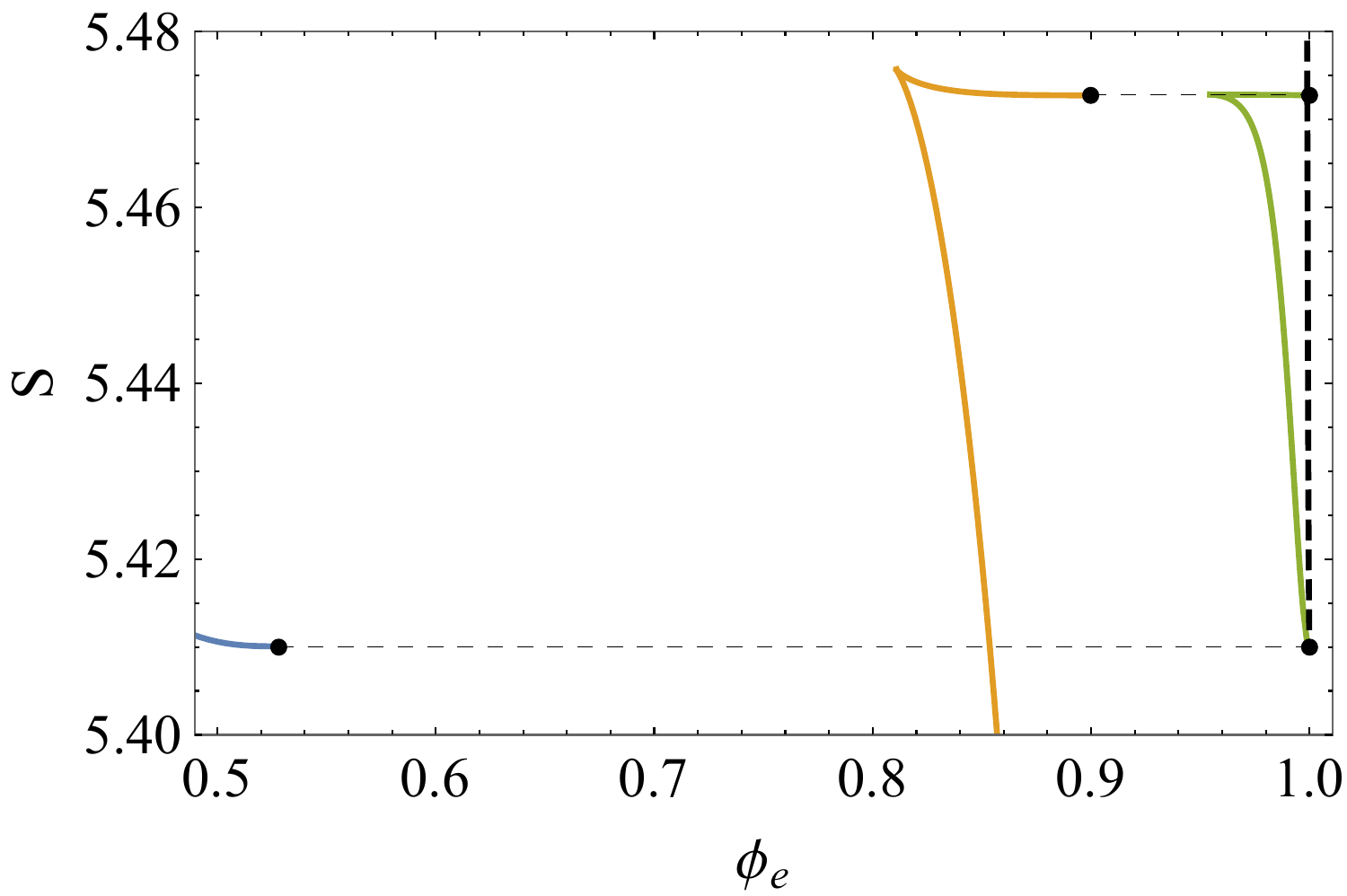}
\end{center}
\caption{Tunneling action for the pseudo-bounce branches of Figure~\ref{fig:ExDri}, with the same color coding. The right plot zooms into the structure of higher $\phi_e$ branches. Black dots are bounces, lying at local minima of $S(\phi_e)$.
\label{fig:ExDS}
}
\end{figure}

Having found the pseudo-bounce solutions, we can also calculate their action, which is given in figure~\ref{fig:ExDS} (with the same color coding used for the $\phi_e$ branches of Figure~\ref{fig:ExDri}). The first family of pseudo-bounces (blue line) has the standard behaviour: the action diverges at a field value $\phi_\infty$
for which $V(\phi_\infty)=V(0)=0$ (dot-dashed line) and then it decreases monotonically towards the bounce value (black dot), with $S\simeq 5.41$. The other pseudo-bounce families have a more interesting behaviour, shown in more detail in the right plot of Figure~\ref{fig:ExDS}. The orange line has a local minimum at $\phi_e=0.9$, as expected by the analytical results, with $S\simeq 5.475$ [in agreement with (\ref{San})]. However, after a kink, the action starts to decrease, as $\phi_e\to 1$, towards $S=0$ (see below). This $S\to 0$ solution is the dominant decay channel in this potential and it is due to the presence of a singularity at $\phi\to 1$, with
\be
V(\phi=1-\delta) = -\frac{1}{6\delta^2}\log^2\phi_0+...
\label{Vexp}
\ee
Apart from this catastrophic decay channel, the green line shows two more bounces also with $\phi_e=1$, one with $S\simeq 5.475$
and the other with $S\simeq 5.41$. Note that these action values are the same as for the analytic and standard solutions respectively, in spite of the fact that these solutions have different profiles (and, in particular, different $\phi_e$). This coincidence (explained below) is indicated by the horizontal thin-dashed lines connecting the two pairs of solutions. Figure~\ref{fig:ExDS} also shows the action for the lowest multi-pass pseudo-bounce, which increases very fast. The action for other higher-pass solutions is even larger.

We can understand some of the features of the tunneling action for pseudo-bounces 
as a function of $\phi_e$ just displayed using the relation 
\be
\frac{dS}{d\phi_e}=\frac{\pi^2}{2}r_i^4(\phi_e) V'(\phi_e)\ ,
\label{dSdphie}
\ee
that was derived in \cite{BoN2} (see Appendix G.1 of that paper). We have numerically checked that (\ref{dSdphie}) holds in this example. The negative slope of the potential at $\phi_e$ gives $S'(\phi_e)<0$ and explains that all branches of the action decrease with increasing $\phi_e$, see Figure~\ref{fig:ExDS}. The slope 
$S'(\phi_e)$ is larger for larger $r_i(\phi_e)$. 
For the standard branch of solutions (blue line) $S(\phi_e)$ decreases towards the stationary point at the bounce (where $r_i=0$ and $S'=0$). In the orange branch, starting from its lowest $\phi_e$ value ($\phi_e\simeq 0.81$), we also understand why the upper part of the $r_i(\phi_e)$ branch, which probes higher values of $r_i$ before dropping to $\phi_e=1$, leads to an action that decreases more steeply compared to the lower  part of the branch, with smaller $r_i$. This implies that the action of the stationary point near $\phi_e=1$ is lower than the action at the stationary point at $\phi_e=0.9$. 
Finally, the green branch, which lives at small values of $r_i$,
has a very flat $S(\phi_e)$ for that reason. A more general discussion of the actions of bounces connected by a pseudo-bounce line is deferred to section~\ref{sect:genless}.

\subsection{Tunneling Potential Approach and $\bma{S(A)}$}
In order to understand better the pseudo-bounce behaviour discussed above it is useful to examine the problem in the light of the tunneling potential formalism. In that language, pseudo-bounce solutions correspond to $V_t$ solutions that leave from the false vacuum and reach $V$ at $\phi_e$ with $V_t'(\phi_e)=0$. Instead of solving the EoM for $V_t$ taking initial conditions at $\phi_e$
and integrating towards $\phi_+=0$, we follow the strategy used in 
\cite{BoN2}: solve the EoM starting from $\phi_+$ to find a full family of pseudo-bounces labelled by a free parameter that appears 
in the low-field expansion of $V_t$. The evolution of the solutions guarantees that pseudo-bounces reach $V$ with the right slope, $V_t'=0$
(except for a discrete number of true bounce solutions with $V_t'=3V'/4$).

\begin{figure}[t!]
\begin{center}
\includegraphics[width=0.6\textwidth]{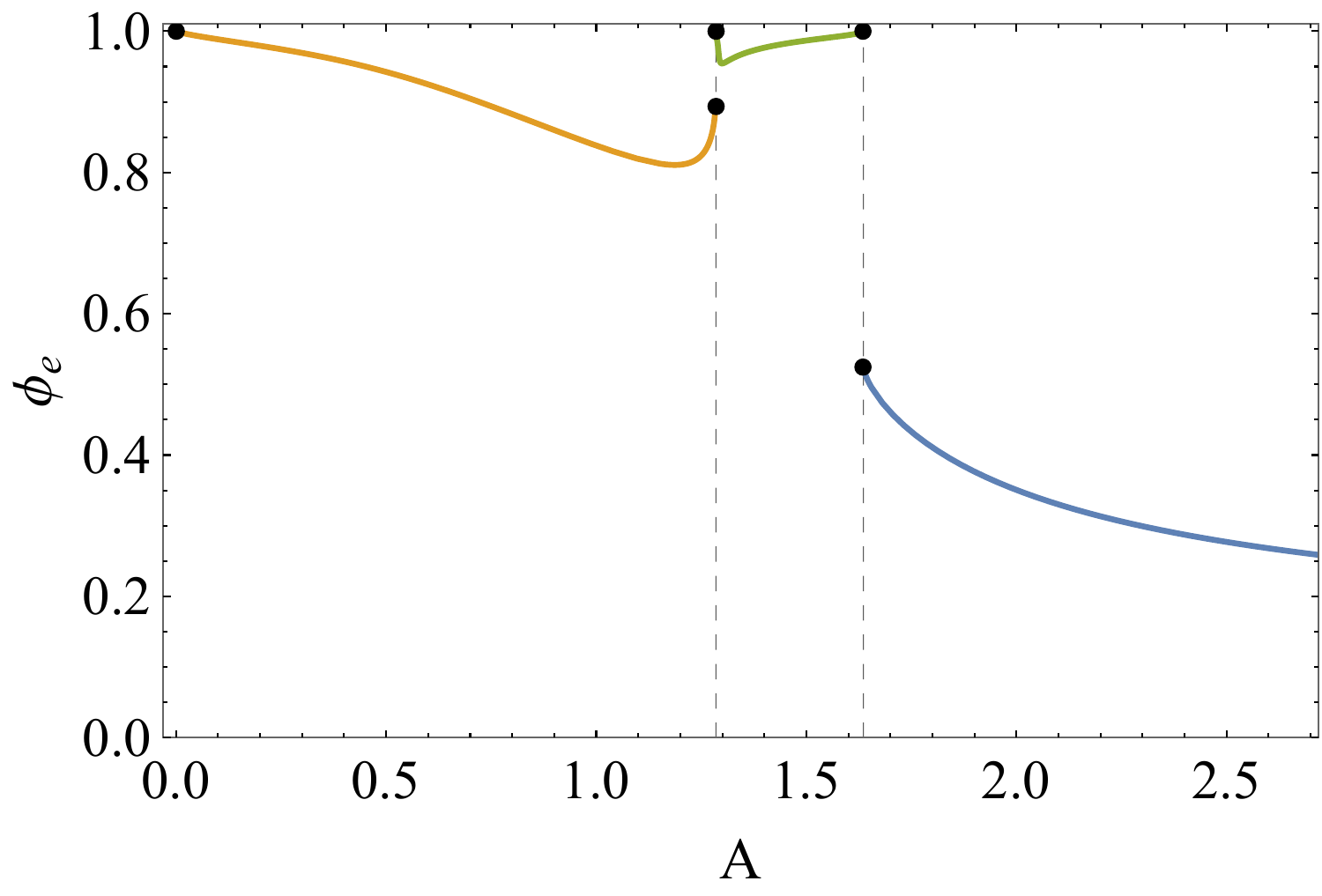}
\end{center}
\caption{End-point $\phi_e$ of the family of tunneling potentials
$V_t(A;\phi)$ for the potential of section~\ref{sect:exD}, using the same color coding of previous figures. The black dots mark the bounce solutions.
\label{fig:ExDVt}
}
\end{figure}

The low-$\phi$ expansions of $V$ and $V_t$ give
\bea
V(\phi) & = & \left[\frac16+\frac{1}{2L_\phi}+\frac{1-2L_0^2/3}{4 L_\phi^2}+\frac{1}{4L_\phi^3}
+{\cal O}(1/L_\phi^4)\right]\phi^2\ ,\\
V_t(A;\phi) & = & \left[\frac{1}{2L_\phi}+\frac{\log A}{L_\phi^2}+
\frac{1/8+2\log^2A}{L_\phi^3}
+{\cal O}(1/L_\phi^4)\right]\phi^2\ ,
\label{VtA}
\eea
where $L_\phi\equiv \log\phi$, $L_0\equiv \log\phi_0$ and $A$ is a free parameter labeling the family of solutions.\footnote{For our numerical analysis we
start the evolution of the $V_t$ solutions at $\phi=10^{-5}$ 
using  (\ref{VtA}) as boundary condition, truncating the expansion at $10^{th}$ order. Including more or less terms changes the numerical value of $A$ associated to a given solution, but $A$ 
is a mere label with no physical meaning.}
Figure~\ref{fig:ExDVt} gives the value of $\phi_e$ as a function of $A$.  Notice that $\phi_e(A)$ is a single-valued function (of course) but some $\phi_e$ values can be reached by two or three values of $A$ and the same $\phi_e$ gap of previous figures is reproduced here.\footnote{The fact that $\phi_e(A)$ is not a monotonic function implies that some $V_t$ pseudo-bounce solutions must cross each other. The possible implications of this fact for the decay action are discussed in Appendix~\ref{app:Vtcross}.}
The thin-dashed black lines mark the two values of $A$ at which
the function $\phi_e(A)$ has discontinuities (see below). The low-$\phi$ expansion of the analytic bounce solution given in eq.~(\ref{Vt09}) shows that $A=e^{1/4}\simeq 1.28$, which agrees with the numerical value shown in  Figure~\ref{fig:ExDVt} for the $\phi_e=0.9$ bounce, at the first discontinuity.

\begin{figure}[t!]
\begin{center}
\includegraphics[width=0.45\textwidth]{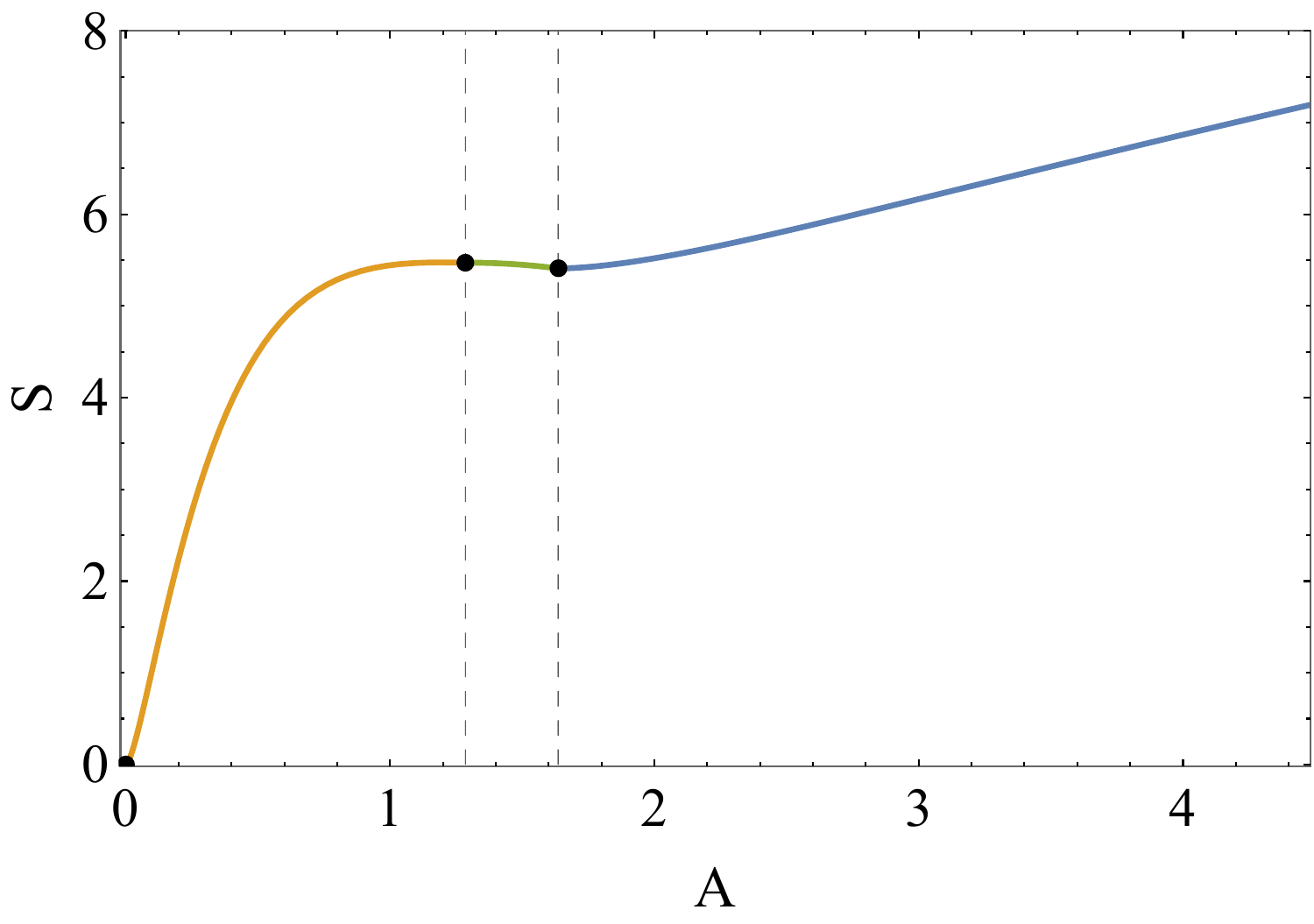}
\includegraphics[width=0.48\textwidth]{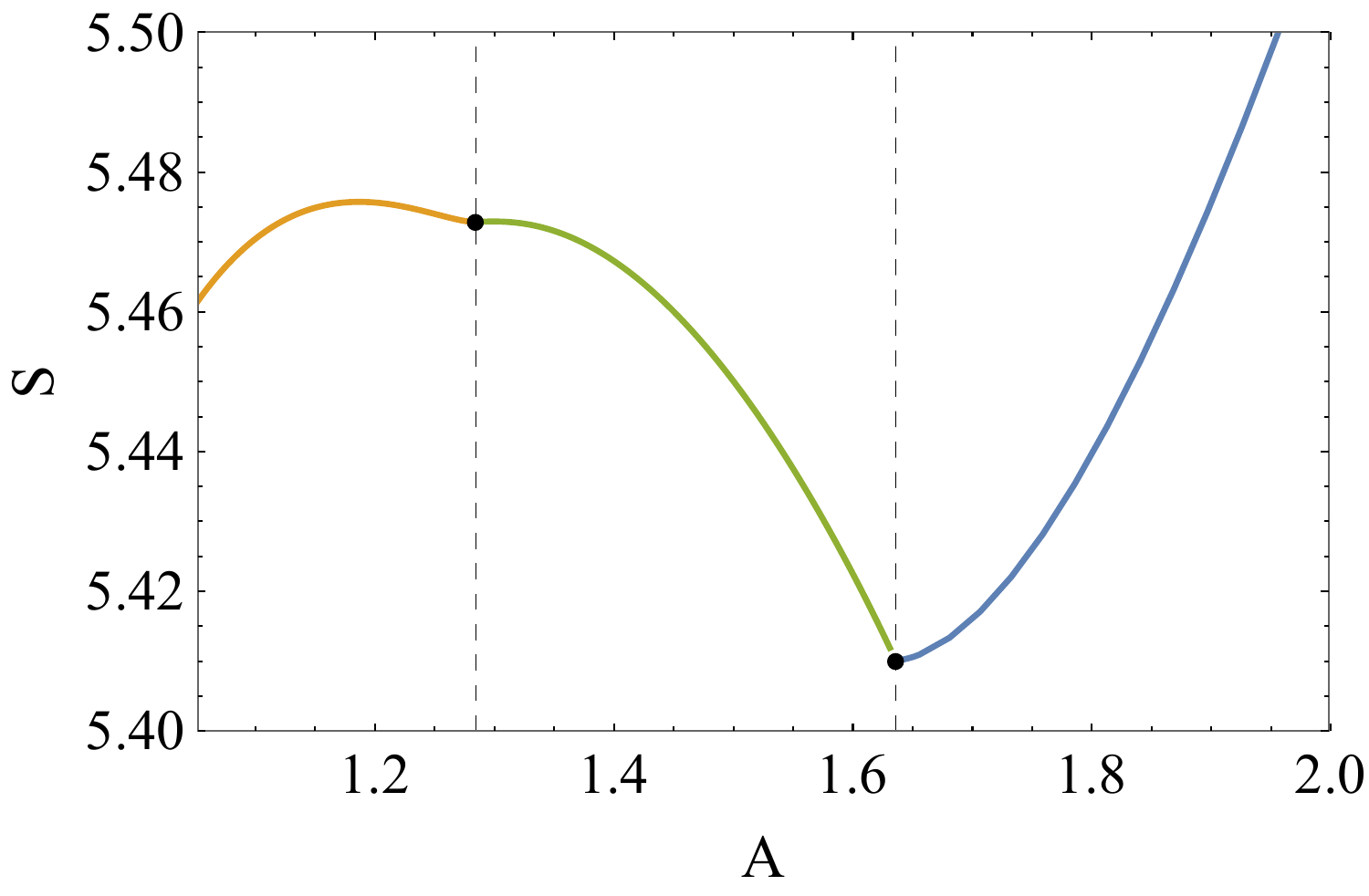}
\end{center}
\caption{Tunneling action for $V_t(A;\phi)$ solutions for the potential of section~\ref{sect:exD}, using the same color coding of previous figures. Dots mark the location of bounces and correspond to local extremals of $S(A)$, and the left plot is just a zoom-in of the right one.
\label{fig:ExDVtS}
}
\end{figure}

\begin{figure}[t!]
\begin{center}
\includegraphics[width=0.45\textwidth]{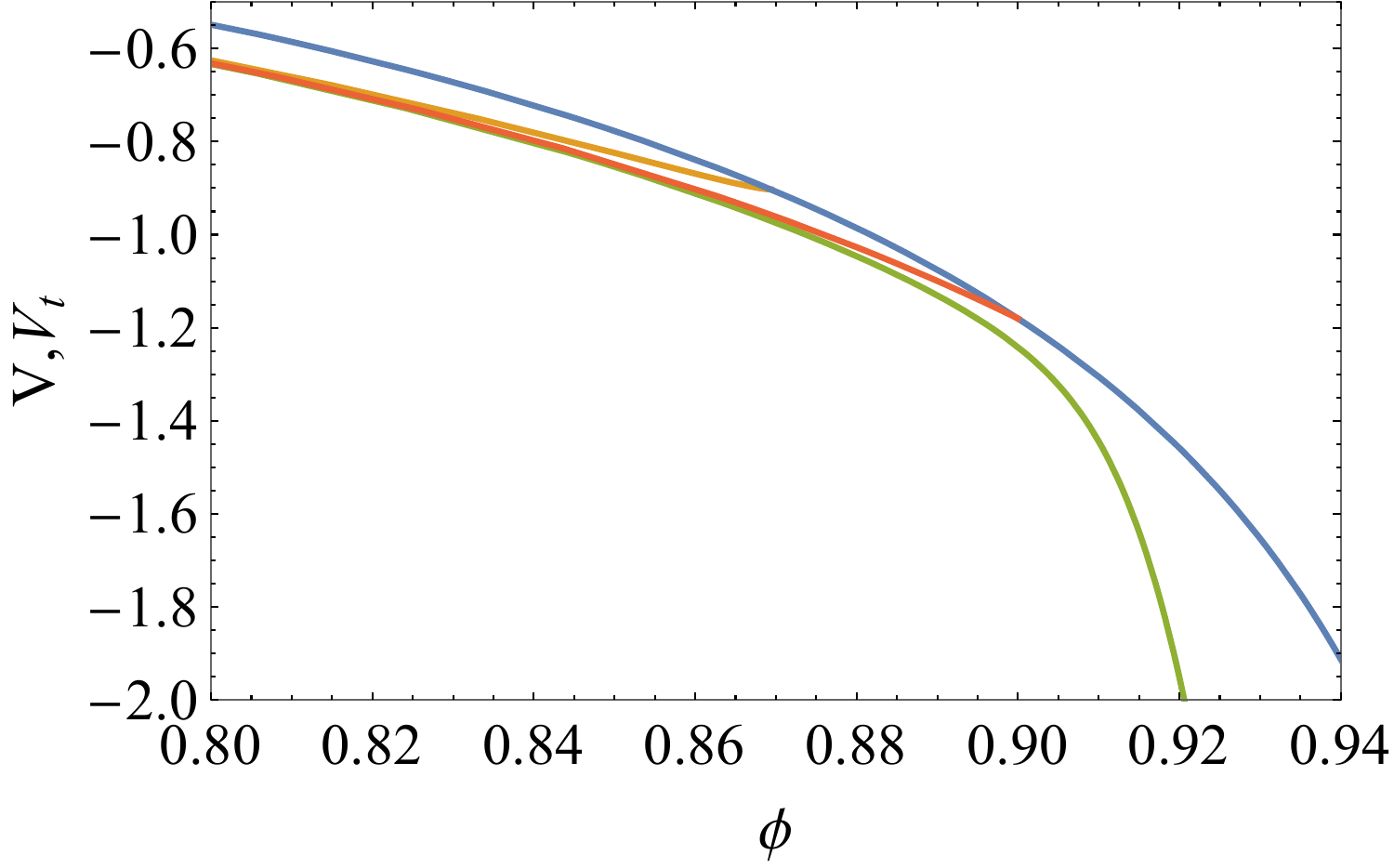}
\includegraphics[width=0.45\textwidth]{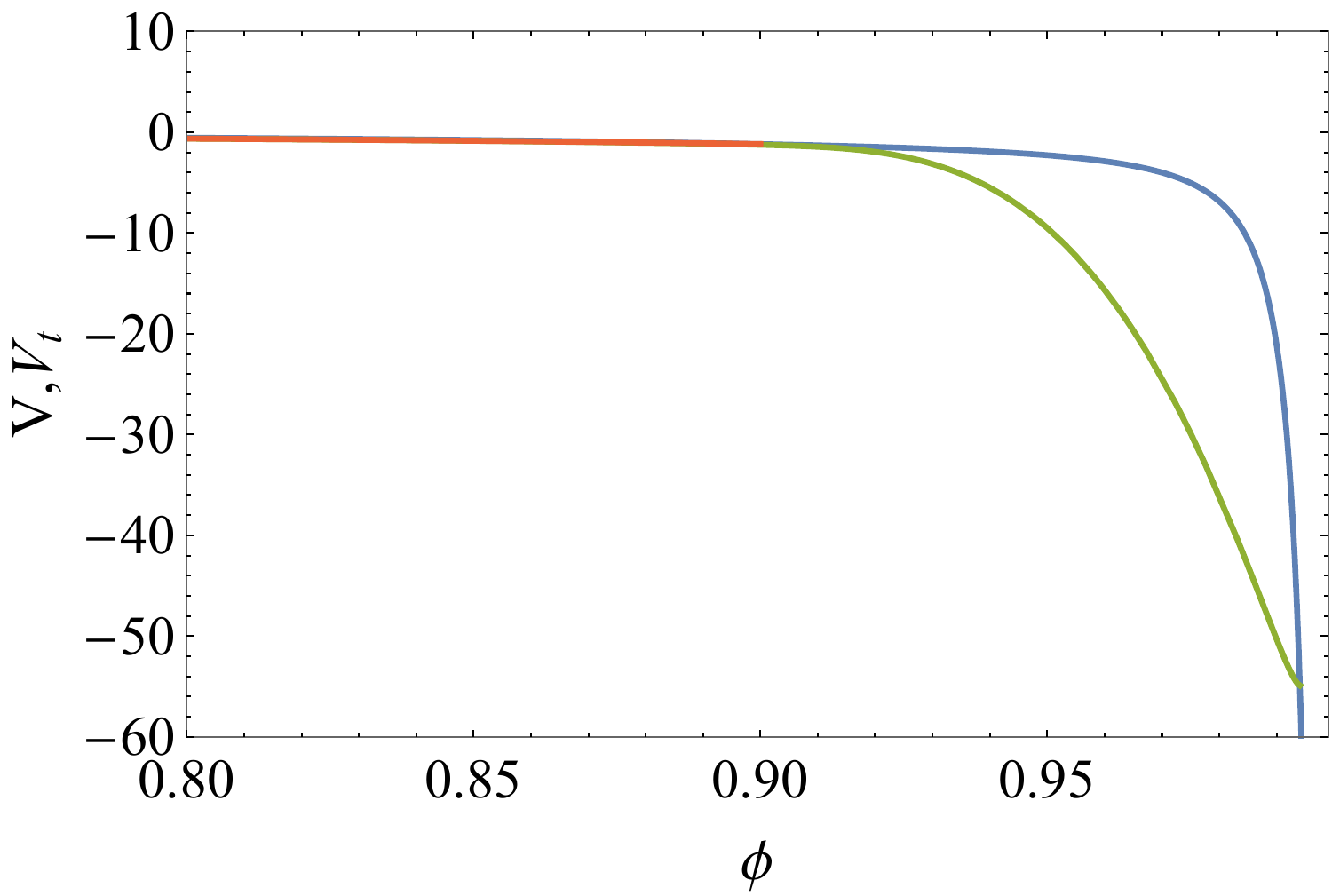}
\end{center}
\caption{Behavior of pseudo-bounce $V_t(A;\phi)$ solutions (orange and green lines) right above and below a true bounce solution (red line) for the potential of section~\ref{sect:exD}. The right plot is a zoomed out version showing the end-point $\phi_e$, where $V_t=V$, for the lower pseudo-bounce.
\label{fig:BelowB}
}
\end{figure}

Interestingly, in spite of the $\phi_e(A)$ discontinuities, the action $S(A)$ is continuous (and single-valued), as shown in Figure~\ref{fig:ExDVtS}. Notice also that, in this parametrization, the lowest action solution goes to zero for $A\to 0$. 
To understand the continuity of $S(A)$ across the jumps in $\phi_e(A)$ let us look more closely at the solutions near these discontinuities.
Figure~\ref{fig:BelowB} shows three $V_t$ solutions near the discontinuity
at $A\simeq 1.28$ (where $\phi_e$ jumps from $\phi_e= 0.9$
to $\phi_e=1$). The left plot shows the (analytic) $V_t$ for the bounce with $\phi_e=0.9$ (red line) and two pseudo-bounce solutions lying close above and below it. The lower solution fails to reach $V$ near $\phi_e=0.9$ and starts to fall very steeply. For a potential with a mild slope beyond $\phi_e$, such solutions would diverge\footnote{With gravity on, such divergent solutions can correspond to bubble of nothing solutions \cite{WittenBoN}, see footnote~\ref{BoNs}.} to $-\infty$.  However, if the potential slope becomes very steep, as in the example, it can catch up with the falling $V_t$ and a new pseudo-bounce solution with finite $\phi_e$ is obtained.
This behaviour of $V$ and $V_t$ is shown in the zoomed out version shown in the right plot of the figure. The story is repeated in the discontinuity at $A\simeq 1.64$ (where $\phi_e$ jumps from $\phi_e= 1$ to $\phi_e\simeq 0.53$). 

The continuity of the action across these jumps follows from the fact that the  action density is very much suppressed by $1/(-V_t')^3$ in the region of steeply falling $V_t$ solutions (while the action densities are nearly identical for lower field values). As these pseudo-bounce solutions tend to the $\phi_e=1$ limit, $V_t'\to -\infty$ and the continuity of $S(A)$ is exact. In the Euclidean approach, the bounces on the two sides of these jumps have the same profile except at their center. The bounces with $\phi_e=1$ have an spike at $r=0$ (reaching up to $\phi_e=1$) and such spike does not give a contribution to the action.

We can also say something about the lowest value of the tunneling action for this potential (which occurs for $A\to 0$ and $\phi_e\to 1$). In that limit, $V$ is very flat compared to $V_t$ and the solution of the EoM for $V_t$, neglecting $V$, gives $V_t\simeq -c \phi^3$.  Using the expansion (\ref{Vexp}) and the boundary condition $V_t(\phi_e)=V(\phi_e)$, we get $c\simeq (\log^2\phi_0)/(6\delta^2)$ and a tunneling action $S\simeq 2\pi^2/c$, which goes to zero as $\delta\to 0$. We have checked numerically that this is a good approximation for the fall-off of $S$ towards zero.

One can contemplate two natural modifications of the previous case: first, one can choose a different $\phi_0$. One finds that the analytic bounce becomes the standard one (blue line branch) for $\phi_0\simlt 0.565$. Otherwise, the overall structure of pseudo-bounce solutions is similar to the one just discussed for 
$\phi=0.9$.\footnote{Concerning the other 1-field analytic examples presented in \cite{EK}, we find normal behavior (just the standard single bounce) in the models that have a true minimum at finite values
(examples A, B and E), while example C (with unbounded potential)
behaves like the example D examined in this section.}
 Second, we can regulate the potential divergence modifying $V(\phi)$ for $\phi>\phi_x$ for some  $\phi_x<1$, {\it e.g.} giving it a finite minimum. This is discussed in section~\ref{sect:exDreg}.

\section{Example from \cite{Sasaki}\label{Sasaki}}
For our second example, we take the potential studied in \cite{Sasaki} (setting here  $\alpha=1, \phi_\star=1, m_1=m_2=1$ for simplicity). It reads 
\be
V(\phi)=\left\{
\begin{matrix}
\frac12 (\phi-\phi_M)^2-\Lambda_2^4&,& \mathrm{for}\; \phi\le \phi_2\\
-\frac12 (\phi-\phi_P)^2-\Lambda_1^4&,& \mathrm{for}\; \phi_2\le \phi\le 0\\
-e^{2\phi} & ,& \mathrm{for}\; \phi\ge 0
\end{matrix}
\right.
\label{VS}
\ee
with parameters chosen so as to have a false minimum at $\phi_M$.
In order to match $V$ and $V'$ at the boundaries ($\phi=0,\phi_2$) between the different field regions in (\ref{VS}) one takes
\be
\phi_P=-2\ , \quad
\phi_2=\frac12 (\phi_M-2)\ , \quad
\Lambda_1^4=-1\ ,\quad
\Lambda_2^4=\frac14 (\phi_M+4)\phi_M\ .
\ee

A singular analytic bounce solution with $\phi_B(0)=\infty$ but with finite action was found in \cite{Sasaki}, where details can be found. We present such bounce solution for our choice of parameters in Appendix~A.\footnote{Similar singular bounces have been found before in the literature as 4d-effective descriptions of higher-dimensional bubble-of-nothing (BoN) solutions, see \cite{BoN0,BoN1,BoN2}. According to the general classification made in \cite{BoN2}, the solutions in \cite{Sasaki} (or Appendix~A) are of the so-called type $-^*$. Notice that, although these BoN singular bounces involve gravity, type $-^*$ ones have an asymptotic behaviour near their singularity (for $\phi\to\infty$) that is independent of gravitational effects \cite{BoN2} and agrees precisely with the no-gravity examples discussed here, which have $V\sim -e^{2\phi}$ and $V_t\sim -(3/2)e^{2\phi}$, see footnote~\ref{Vtfactor}.
\label{BoNs}} 
For the analytic bounce to satisfy the correct boundary conditions, the value of $\phi_M$ has to be tuned. With our choice of parameters above we get $\phi_M\simeq -2.36$ and this fixes all the free parameters of the potential. The action of the analytical bounce is then $S_0\simeq 130.44$, see Appendix~A for the details.

\begin{figure}[t!]
\begin{center}
\includegraphics[width=0.475\textwidth]{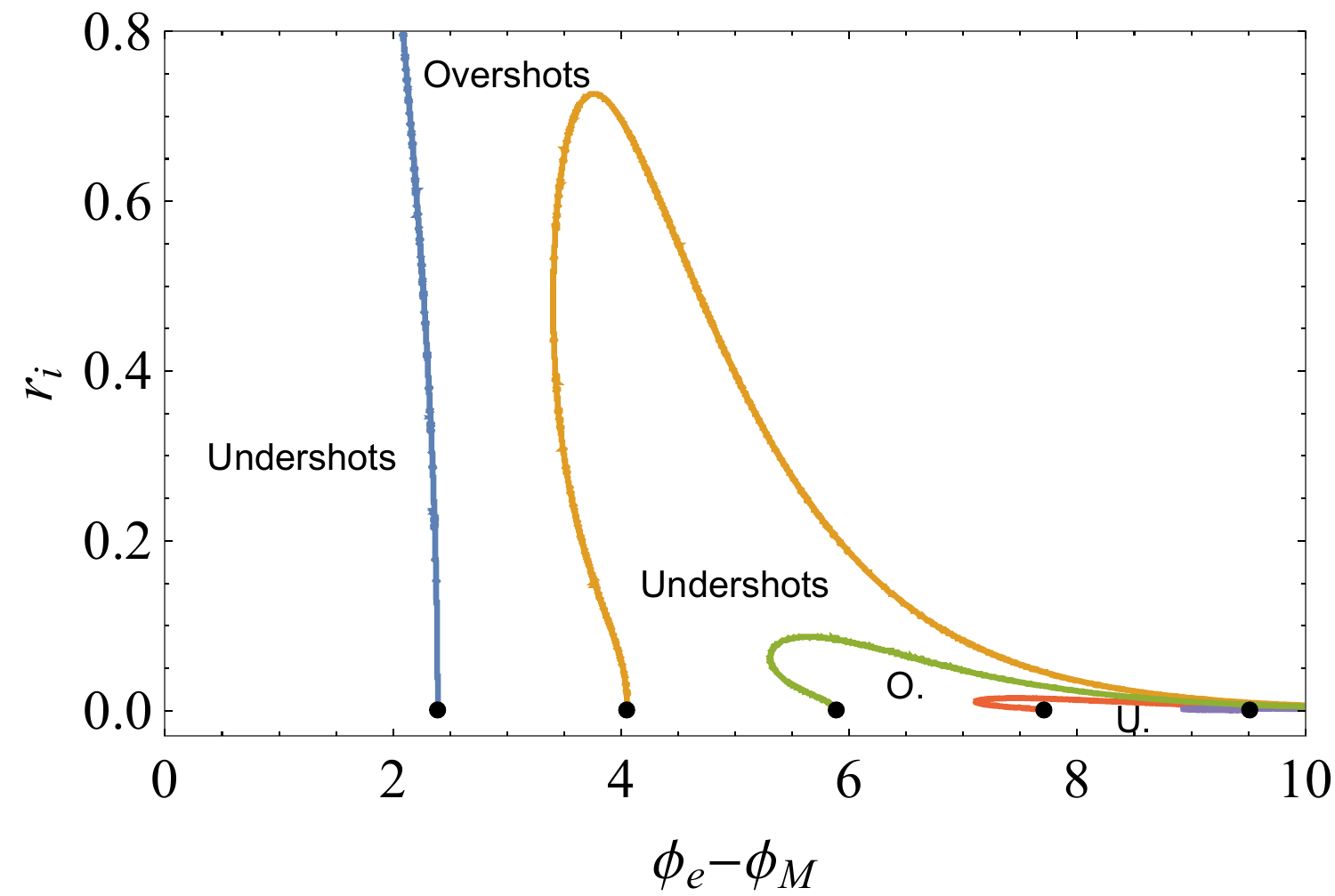}\;
\includegraphics[width=0.475\textwidth]{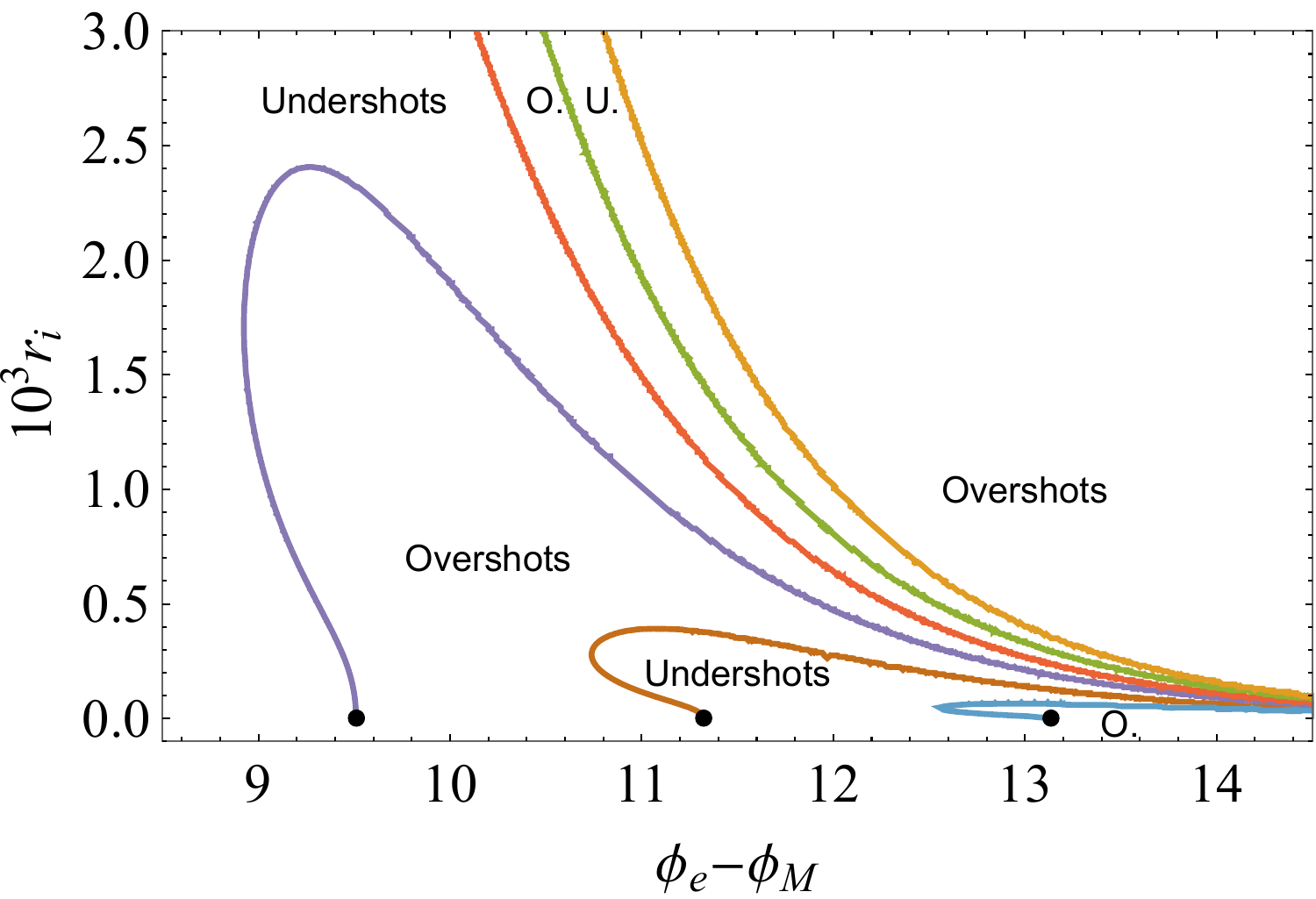}
\end{center}
\caption{For the example of section~\ref{Sasaki}, different pseudo-bounce branches (with  core field value $\phi_e$ and  inner core radius $r_i$) separating overshot and undershot regions. Bounces live along the $r_i=0$ axis and are indicated by black dots. The right plot zooms to higher values of $\phi_e$. The same structure continues even further all the way to $\phi_e\to \infty$ and $r_i\to 0$.
\label{fig:Sasakiphieri}
}
\end{figure}

We can repeat the analysis of the previous section for this potential.
The pseudo-bounce lines in the ($\phi_e, r_i$) plane are shown in Figure~\ref{fig:Sasakiphieri}.\footnote{We do not assume anything about the behavior of $V$ to the left of the false vacuum and thus we do not discuss here multi-pass pseudo-bounces, see footnote~\ref{Vsym}.} We find a series of bounces (marked by black dots in the figure) with $\phi_e-\phi_M=\{2.39,4.05,5.88,7.70,9.51,11.32,13.14,14.95,...\}$ which continues indefinitely, bounce alternating with antibounce, with an spacing that tends to a constant $\Delta\phi_e\simeq 1.81$. In the figure, the right plot shows how the structure of pseudo-bounces shown in the left plot continues to higher $\phi_e$, with the $r_i$ values probed being progressively smaller. An explanation for this pattern is given below. 

As we learned in the previous section, it is convenient to switch to the tunneling potential approach to get a continuous tunneling action. The $V_t$ approach is also useful in this example to understand the infinite series of true bounces discussed above, as we discuss later. The parameter $A$ to describe the $V_t$ solutions appears as before in the expansion of solutions near the false vacuum. For the potential, for $\phi\sim\phi_M$, we have
\be
V(\phi)=-\Lambda_2^4+\frac12 (\phi-\phi_M)^2\ ,
\ee
while for $V_t$ we get
\be
V_t(A;\phi)=-\Lambda_2^4+(\phi-\phi_M)^2\left[\frac{1}{W}+\frac{2}{3W^2}-\frac{1}{9W^3}+\frac{5}{72W^5}+{\cal O}(1/W^6)\right]\ ,
\label{VtW}
\ee
where $W\equiv {\rm ProductLog}[A(\phi-\phi_M)^{-2/3}]$, with the product log\footnote{To see the origin of $W$, one can first derive the expansion (for $r\to \infty$) of the Euclidean pseudo-bounces: $\phi(r)-\phi_M=C K_1(r)/r = r^{-3/2} e^{-r} [1 + O(1/r)]$, where $K_1$ is the Bessel function of the second kind and $C$ an arbitrary constant. This leads to  $r \simeq (3/2) W[A(\phi-\phi_M)^{-2/3}]$ with
$A$ related to $C$ by $A=(2\sqrt{\pi}C)^{2/3}/3$. The expansion of $V_t$ in (\ref{VtW}) follows from $V_t=V-\dot\phi^2/2$. For $\phi\to\phi_M$ one can further expand $W(x)$ at large $x$ as $W(x)=\log x+(1-\log x)\log(\log x)/(\log x)+...$} satisfying $W(x) e^{W(x)}=x$. We use this analytic expression for the boundary conditions used in the numerical analysis.

In the exponential region of the potential ($\phi>0$), the behavior of $V_t$ is quite simple. If a solution $V_t(\phi)$ is known in that part of the potential, additional 
solutions can be generated via\footnote{In the Euclidean approach one has $\tilde \phi(r) = \phi(r \, e^{\delta \phi})   + \delta \phi$. This relation is behind the repeated pattern of solutions in Figure~\ref{fig:Sasakiphieri}.}
\be
\tilde V_t(\phi) = e^{-2\delta \phi} \, V_t(\phi + \delta \phi)\, .
\ee
This follows from the form of the equation of motion (\ref{EoMVt}) and the fact that the exponential potential fulfills $V(\phi + \delta \phi) = V(\phi) \, e^{2\delta\phi}$.

\begin{figure}[t!]
\begin{center}
\includegraphics[width=0.45\textwidth]{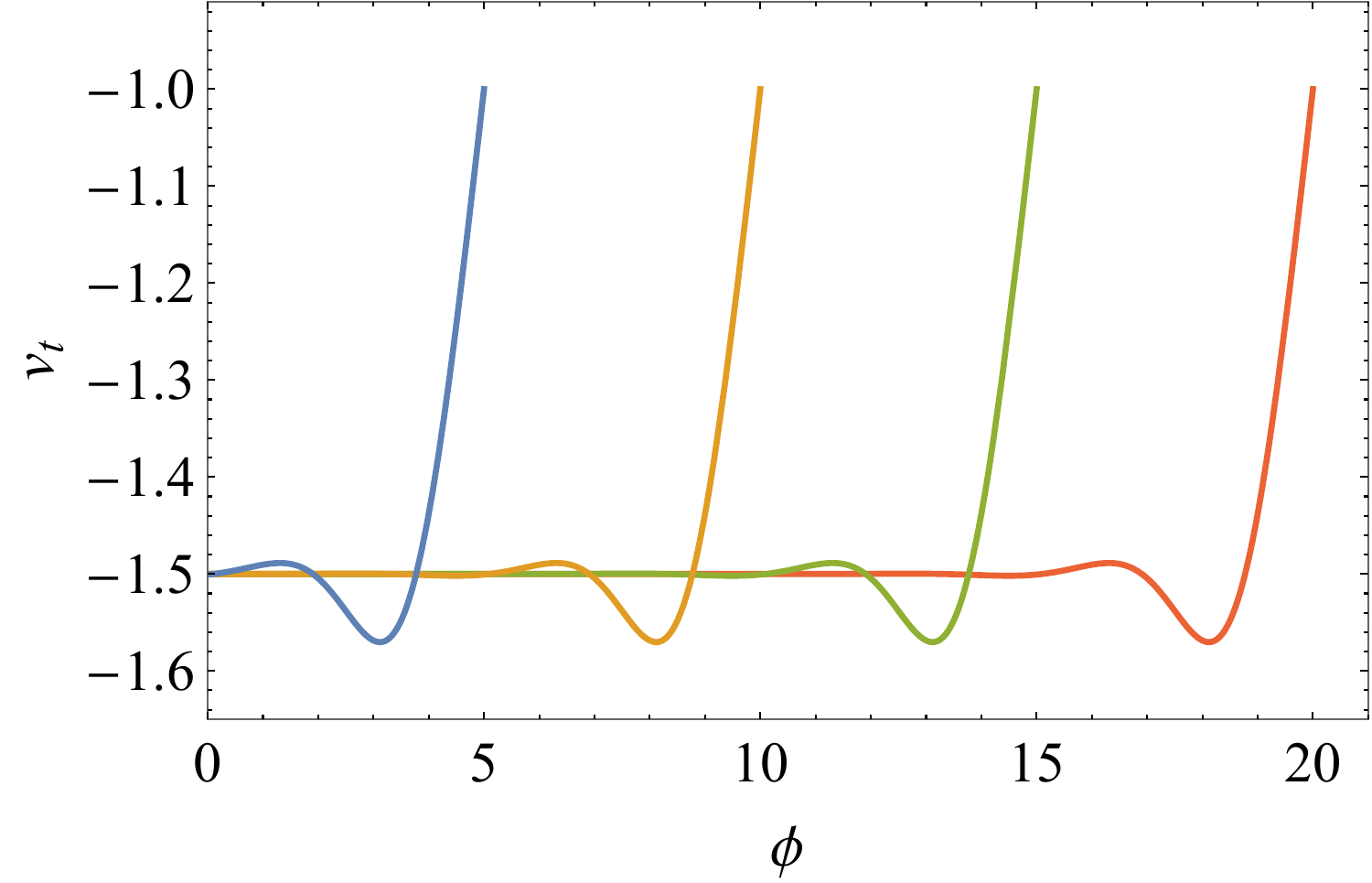}\;\;
\includegraphics[width=0.45\textwidth]{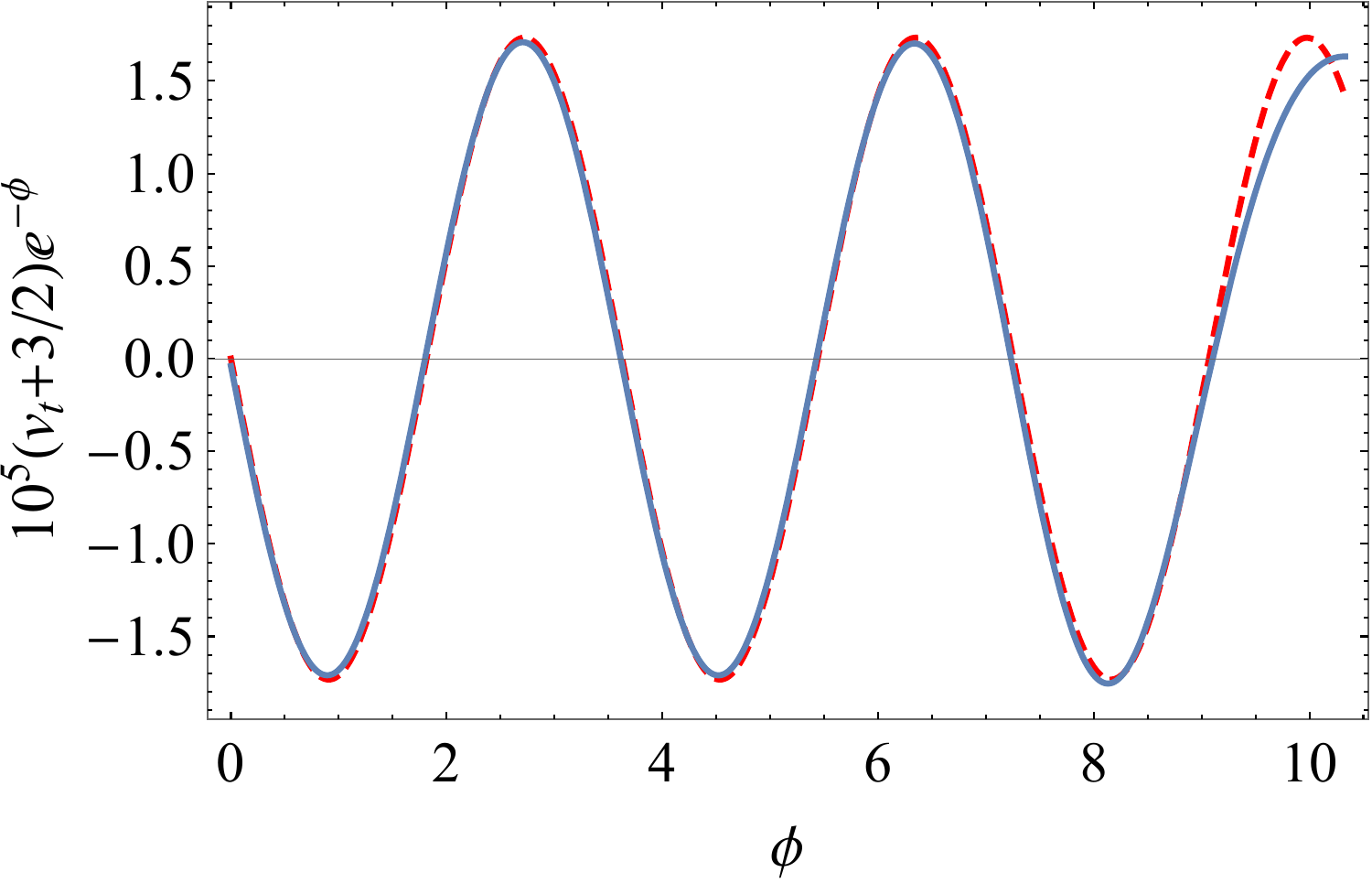}
\end{center}
\caption{Left: Different $v_t(\phi)$ solutions of (\ref{EoMvt}) with boundary conditions (\ref{vtBCs}) and different values of $\phi_e=\{5,10,15,20\}$ are simply shifted copies of each other. Right: Example of the oscillatory behavior (\ref{vtapp}), for $\phi_e=10.33$.
\label{fig:vt}
}
\end{figure}

Alternatively, one can write
\be
V_t(\phi)=v_t(\phi)e^{2\phi}\ ,
\ee
and plugging this in (\ref{EoMVt}), the equation of motion for $v_t$ is obtained as
\be
(4v_t+6)v_t+(9+4v_t-2v_t')v_t'+3(1+v_t)v_t''=0\ .
\label{EoMvt}
\ee
This equation has no explicit dependence on $\phi$ and thus 
any solution $v_t(\phi)$ can be shifted in $\phi$ at will.
The boundary conditions at $\phi_e$ for a bounce ($V_t=V$ and $V_t'=3V'/4$) give
\be
v_t(\phi_e)=-1\ ,\quad v_t'(\phi_e)=1/2\ .
\label{vtBCs}
\ee
These simple boundary conditions are independent of $\phi_e$, and this implies that the  solutions satisfying these boundary conditions must be functions of $\phi-\phi_e$. Therefore
$v_t(\phi)$ solutions with different $\phi_e$ are just 
trivially related by a constant shift in $\phi$. Figure~\ref{fig:vt}, left plot,  shows several such solutions for $\phi_e=\{5,10,15,20\}$. At $\phi\to 0$ (the lower boundary of the exponential region) the solutions are close to the asymptotic value $v_t=-3/2$, which is
an exact solution for $\phi_e\to\infty$ and corresponds to the singular bounce discussed in Appendix~A.\footnote{Indeed, using $V_t=V-\dot\phi_B^2/2$ and $\phi_B=-\log r$ [see (\ref{phiexp})], one gets $V_t=-(3/2)e^{2\phi}$. \label{Vtfactor}} 

For finite $\phi_e$, the value $v_t=-3/2$ is approached in a damped oscillatory manner and, for small $\phi$, this oscillatory behavior can be well approximated analytically. Without loss of generality (thanks to the shift symmetry discussed above), consider the solution with $v_t(0)=0$ 
and large $\phi_e$ so that $v_t'(0)\equiv v_{tp}$ is small. An expansion in powers of $v_{tp}$ gives
\be
v_t(\phi)= -\frac32 + \frac{v_{tp}}{\sqrt{3}} e^\phi\sin(\sqrt{3}\phi)+{\cal O}(v_{tp}^2)\ .
\label{vtapp}
\ee
Figure~\ref{fig:vt}, right plot, shows this oscillatory behavior both for a numerical solution (continuous blue line) and for the approximation (\ref{vtapp}) (dashed red line). We plot the quantity $(v_t+3/2)e^{-\phi}$ to isolate the pure oscillatory component of the solution. We have taken $\phi_e=10.33$, adjusted to produce $v_t(0)=0$.

\begin{figure}[t!]
\begin{center}
\includegraphics[width=0.45\textwidth]{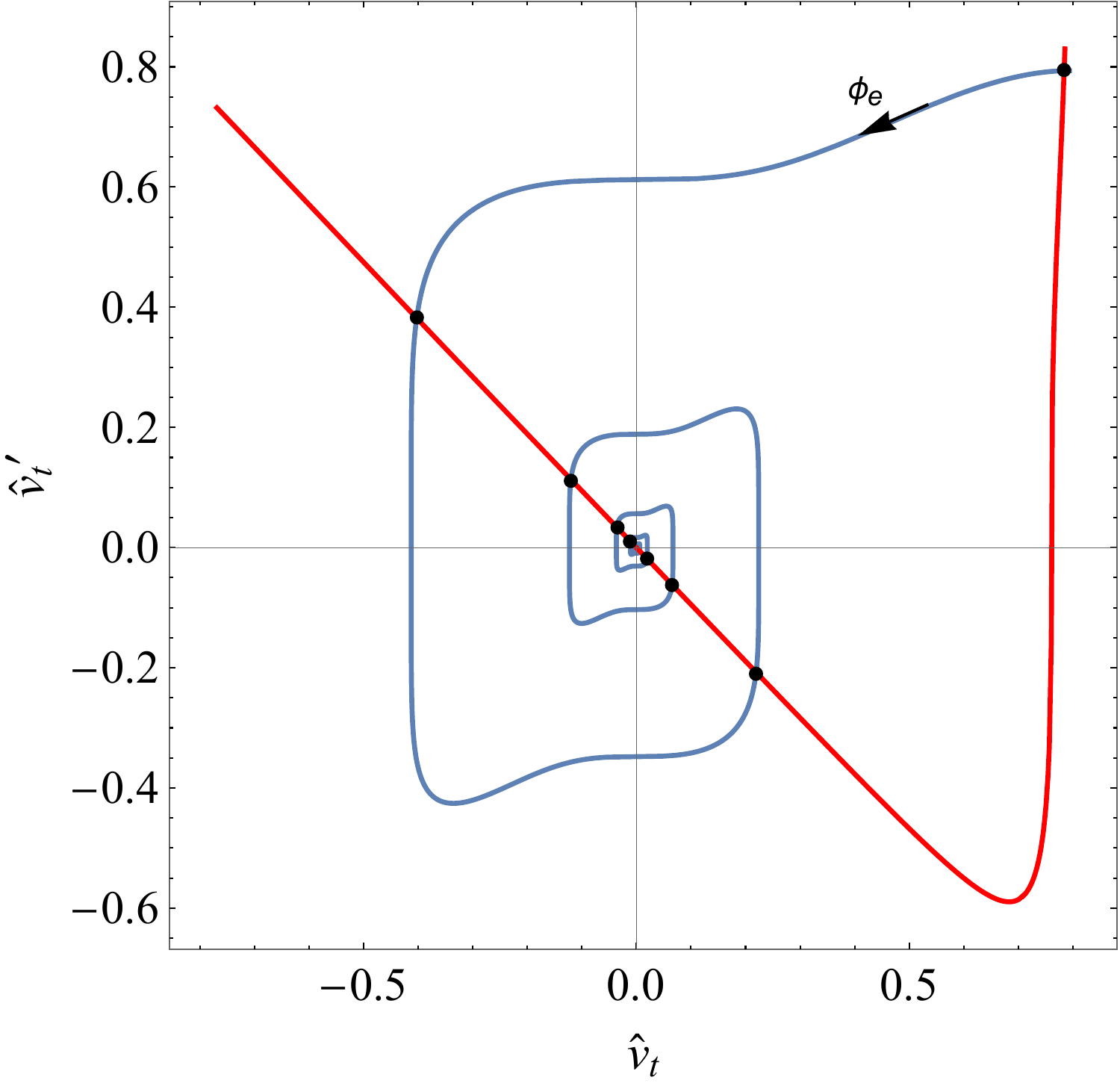}
\end{center}
\caption{Blue spiral line: values of $\hat{v}_t,\hat{v}'_t$, defined in (\ref{vthat}), for solutions from $\phi_e$ down to $\phi=0^+$ (with $\phi_e$ increasing as indicated by the arrow). Red line: same for solutions from $\phi_M$ to $\phi=0^-$. Intersections, marked by black dots, indicate the bounce solutions that satisfy the proper matching at $\phi=0$.
\label{fig:spiral}
}
\end{figure}

\begin{figure}[t!]
\begin{center}
\includegraphics[width=0.45\textwidth]{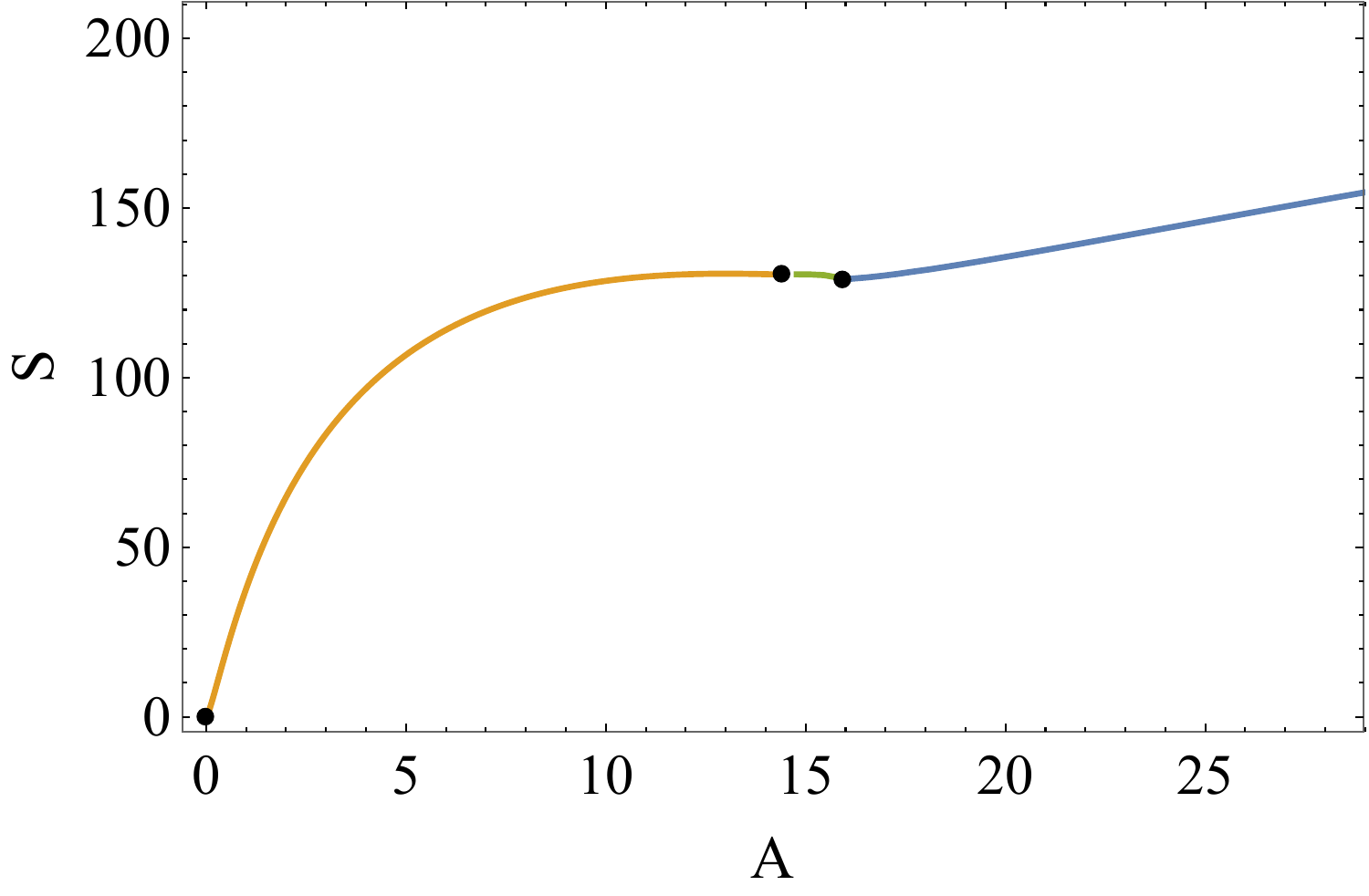}\;\;
\includegraphics[width=0.45\textwidth]{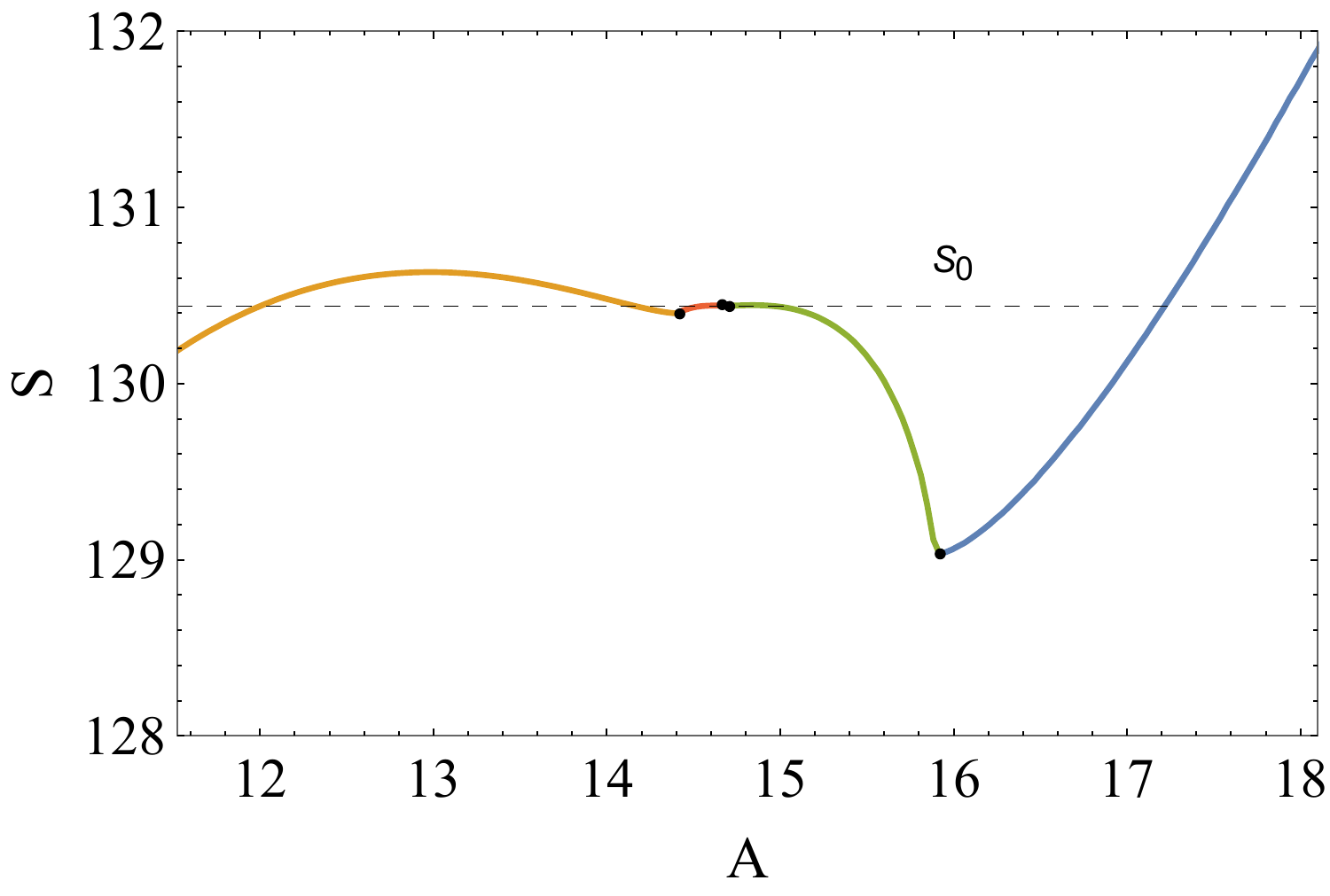}
\end{center}
\caption{Tunneling actions for $V_t(A;\phi)$ solutions for the example considered in this section, using the same color coding of Figure~\ref{fig:Sasakiphieri}. Dots mark the location of bounces [local extremals
of $S(A)$] with an accumulation point around $A\simeq 14.7$, as shown in the zoomed-in right plot. 
\label{fig:SasakiS}
}
\end{figure}

\begin{figure}[t!]
\begin{center}
\includegraphics[width=0.45\textwidth]{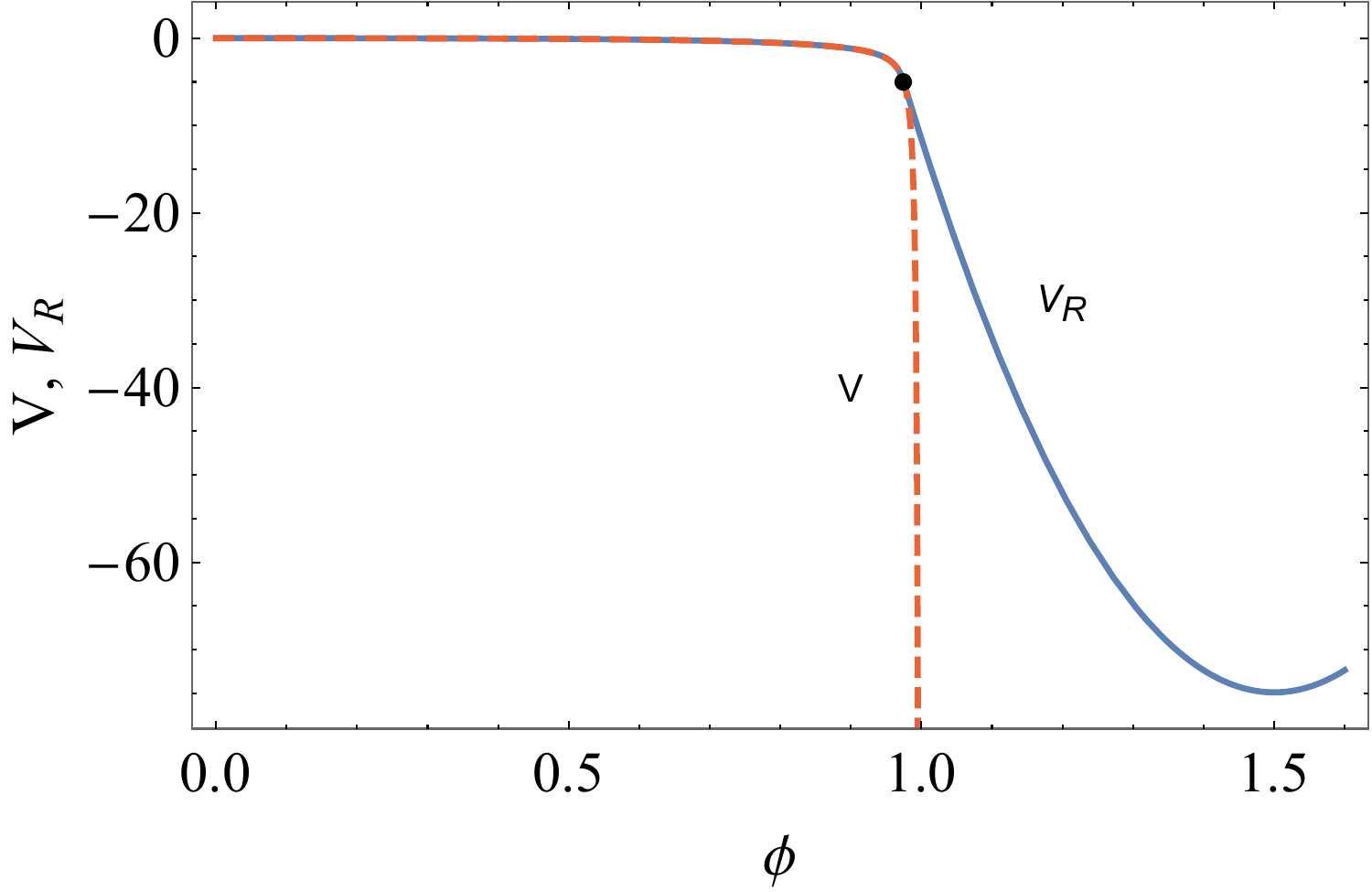}
\end{center}
\caption{
\label{fig:Vreg} Regularized potential (\ref{VDreg}), $V_R(\phi)$, (solid blue) and unbounded-from-below potential $V(\phi)$ of section~\ref{sect:exD} (red dashed). The matching $\phi_x$ is marked by a black dot.
}
\end{figure}  

To be proper bounces, the $v_t(\phi)$ solutions just discussed should arrive at $\phi=0$ with the right values of $v_t(0^+)$ and $v_t'(0^+)$  to reach the false vacuum as $\phi\to\phi_M$. Such correct values can be obtained by numerically solving for $V_t$ in the field interval $(\phi_M,0^-)$. As discussed above, there is an infinite family of such solutions, parametrized by $A$ in the expansion (\ref{VtW}). This procedure gives a curve in the plane $(V_t,V_t')$ [or equivalently $(v_t+3/2,v_t')$] at $\phi=0^-$. The intersection(s) of that curve with the corresponding curve for $\phi=0^+$ (from the $v_t$ solutions in the exponential region)  give the different
bounces (satisfying the proper matching at $\phi=0$). Due to the oscillatory nature of the $v_t$ solution in the exponential region, the latter curve is an spiral with center at 
$v_t=-3/2$ and $v_t'=0$ (the singular bounce). In order to see better the spiral structure near this center we plot in Figure~\ref{fig:spiral} the quantities
\be
\hat{v}_t\equiv {\rm sign}(v_t(0)+3/2)\ |v_t(0)+3/2|^{1/3}\ ,\quad\quad
\hat{v}'_t\equiv {\rm sign}(v'_t(0))\ |v'_t(0)|^{1/3}\ .
\label{vthat}
\ee
The spiral (blue curve) corresponds to  $\phi=0^+$ and the red curve to $\phi=0^-$. Along the spiral, the value of $\phi_e$ increases as indicated by the arrow, with the center corresponding to $\phi_e\to\infty$. We see that there is an infinite series of intersections and thus of bounce solutions, with increasing values of $\phi_e$. In fact, from the approximation (\ref{vtapp}), the jump in $\phi_e$ from solution to solution tends
to $\Delta\phi_e=\pi/\sqrt{3}\simeq 1.81$, in agreement with the numerical result obtained previously in the Euclidean approach. Note that the existence of an infinite number of bounces follows from the fact that the red line crosses the spiral center, which is guaranteed by construction: the potential parameters have been tuned in order to have a singular solution.

As a function of the $A$ parameter, the tunneling action is shown in Figure~\ref{fig:SasakiS}.
As in the previous example, the action is a continuous
function of $A$. The infinite series of bounces discussed above accumulate in the region where $S(A)$ flattens, around
$A\simeq 14.7$, with actions that tend to $S_0=130.44$, the value for the singular bounce (shown with a dashed line). 
We nevertheless see that there are solutions 
with lower action, and in fact $S(A)\to 0$ for $A\to 0$ as in the example of the previous section and with a similar explanation in terms of a cubic $V_t\simeq -C(\phi-\phi_M)^3$.

\section{Example from \cite{EK} Regularized\label{sect:exDreg}}
The scalar potentials of sections~\ref{sect:exD} and~\ref{Sasaki} are unbounded from below. The one of section~\ref{sect:exD} diverges to $-\infty$ at a finite field value ($\phi=1$) while the one of section~\ref{Sasaki} diverges like $V(\phi)=-e^{2\phi}$ for $\phi\to\infty$. In both cases the false vacuum is badly unstable, with tunneling action $S\to 0$.
In this section we explicitly show that the type of behaviour found in those sections (in particular the existence of additional bounces and antibounces) does not hinge on such pathological properties of the potentials. To do this it is enough to consider the potential of section~\ref{sect:exD} modified by giving it a true minimum with a finite potential value.

The regularized potential is the following:
\be
V_R(\phi)=\left\{
\begin{matrix}
V(\phi)
&& \mathrm{for}\; 0\le \phi\le \phi_x\\
V(\phi_x)+\frac12 V'(\phi_x)\left[(\phi_--\phi_x)-\frac{(\phi-\phi_-)^2}{(\phi_--\phi_x)}\right]& & \mathrm{for}\; \phi\ge \phi_x
\end{matrix}
\right.
\label{VDreg}
\ee
where $\phi_x<1$, and $V(\phi)$ is the potential in (\ref{VexD}). In other words, below $\phi_x$ the potential is the same as in section~\ref{sect:exD} and above $\phi_x$ the potential has a minimum at $\phi_-$. This regularized potential is constructed to have $V_R$ and $V_R'$ continuous at $\phi_x$. For the numerical analysis below we chose $\phi_0=0.9$
(as in section~\ref{sect:exD}), $\phi_x=0.975$ and $\phi_-=1.5$. The potentials $V$ and $V_R$ for this choice of parameters are shown in Figure~\ref{fig:Vreg}.

\begin{figure}[t!]
\begin{center}
\includegraphics[width=0.5\textwidth]{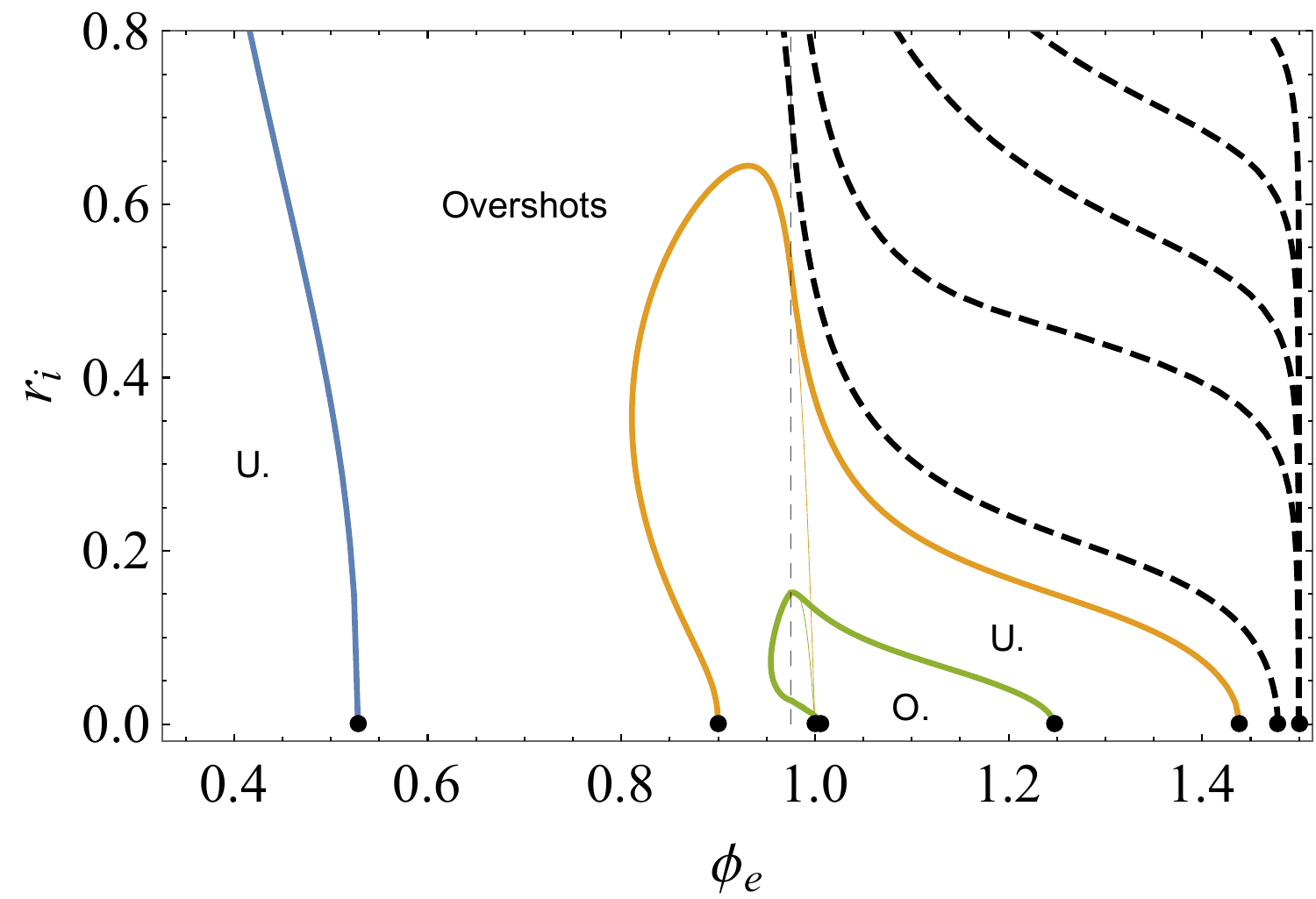}
\end{center}
\caption{As for Figure~\ref{fig:ExDri} but for the regularized potential (\ref{VDreg}). Thick lines correspond to the regularized case and thin ones to the original potential. The matching field value $\phi_x=0.975$ is given by the dashed vertical line. Overshot and undershot regions are labelled accordingly and bounces are marked by black points.
\label{fig:ExDri_reg}
}
\end{figure}

Figure~\ref{fig:ExDri_reg} shows the pseudo-bounce curves in the $(\phi_e,r_i)$ plane. The thin curves above $\phi_x$ correspond to the original potential, precisely as in Figure~\ref{fig:ExDri} (with the same color coding and omitting multi-pass solutions to avoid clutter). The thick lines correspond to the regularized potential. Below $\phi_x$ (marked by the vertical dashed line) the curves are not modified, of course, as pseudo-bounces below $\phi_x$ cannot be sensitive to changes in the potential at $\phi>\phi_x$. The pseudo-bounce curves above $\phi_x$ are deformed and the three bounces with degenerate value $\phi_e=1$ are now split, with $\phi_e\simeq\{1.005,1.247,1.439\}$. The plot also shows the multi-pass solutions for $V_R$, assuming $V_R(-\phi)=V_R(\phi)$.

\begin{figure}[t!]
\begin{center}
\includegraphics[width=0.45\textwidth]{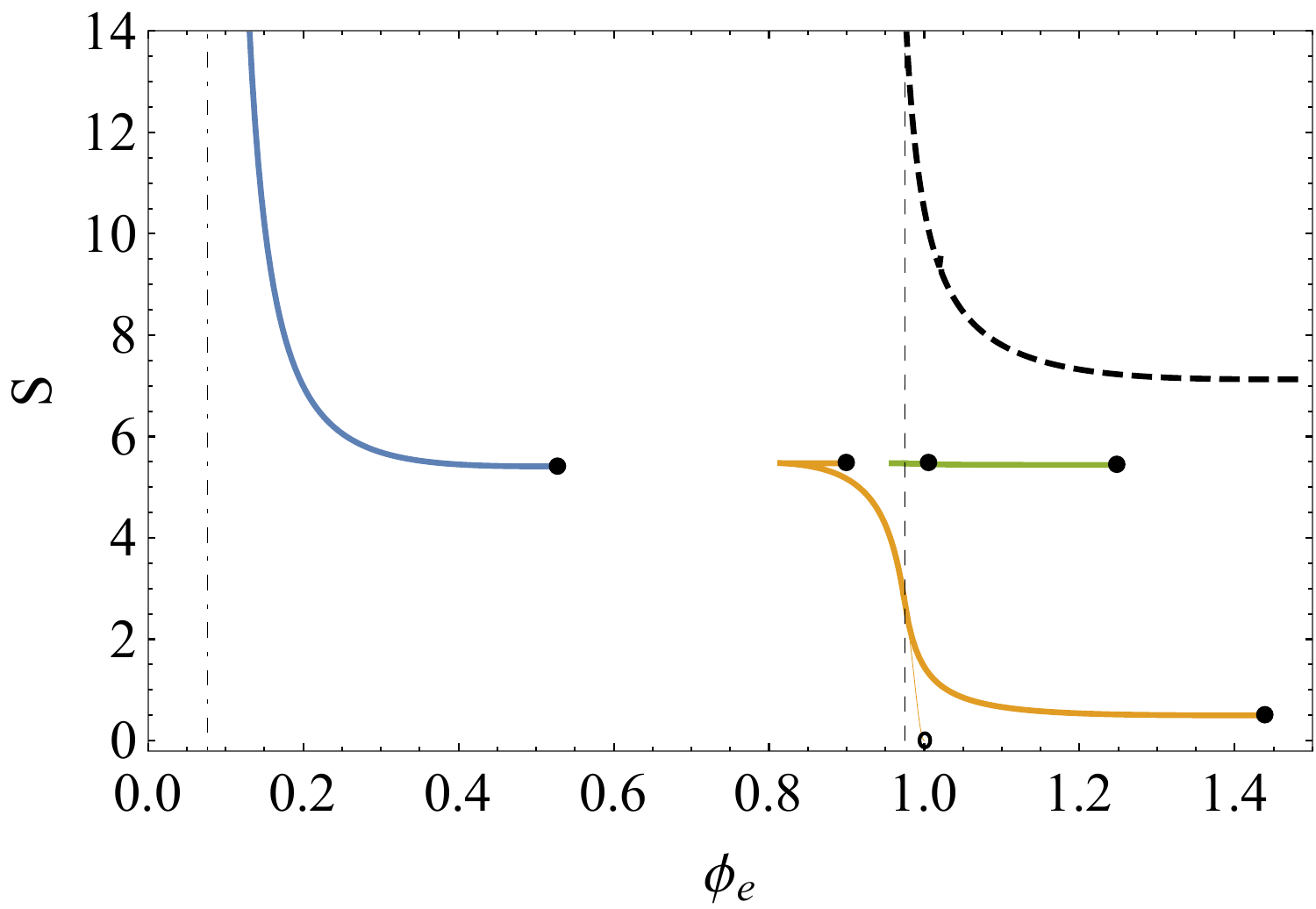}
\includegraphics[width=0.47\textwidth]{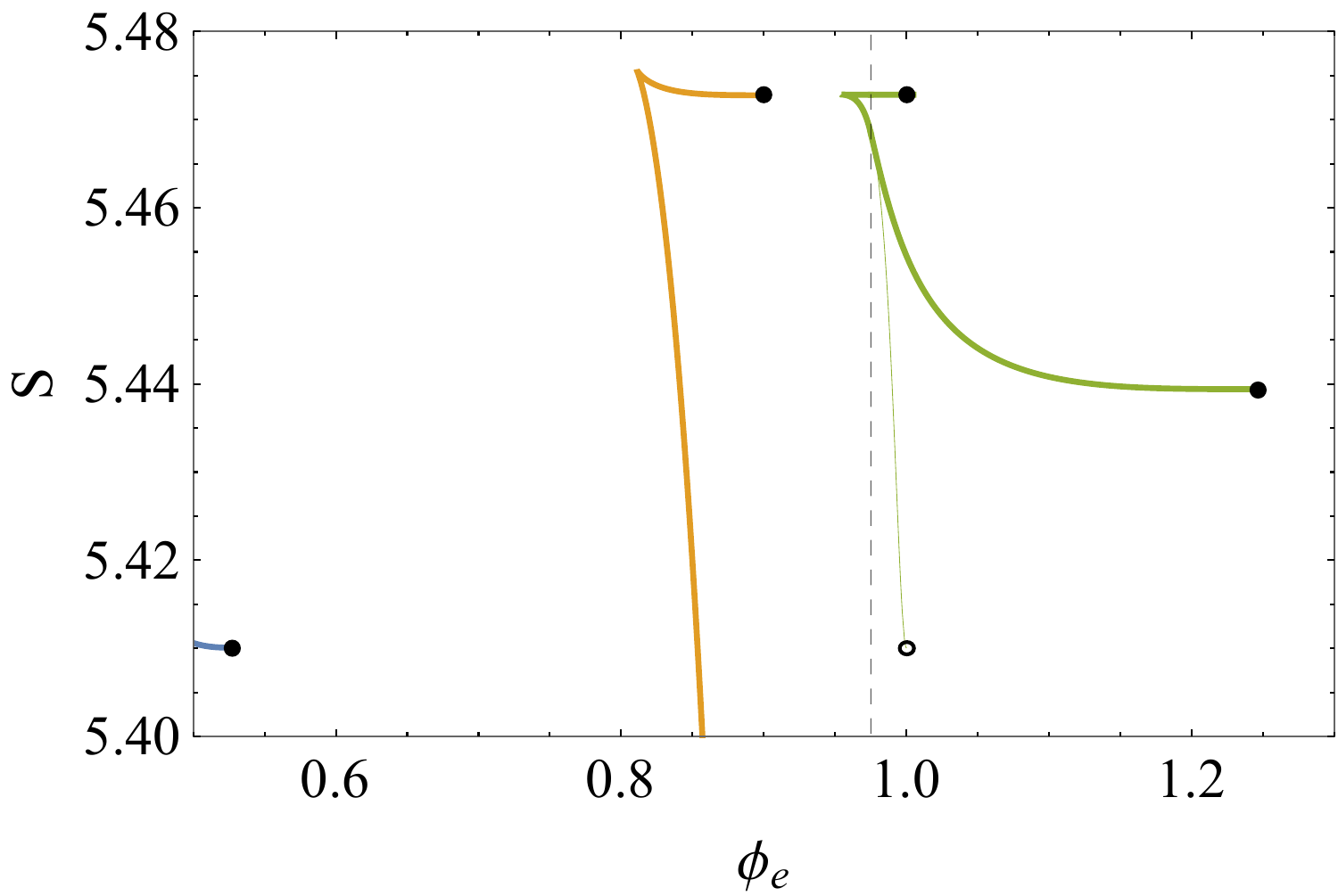}
\end{center}
\caption{Same as Figure~\ref{fig:ExDS} but for the regularized potential of this section.
Thin lines correspond to the original potential $V$, as in Figure~\ref{fig:ExDS}. Empty-circle points indicate bounces for $V$ and black points for $V_R$.
\label{fig:ExDS_reg}
}
\end{figure}

The tunneling action $S(\phi_e)$ is shown in Figure~\ref{fig:ExDS_reg}, which also gives  the action for the unbounded potential $V$ (thin lines and empty-circle points for bounces). The zoomed-in version of the plot
shows how the action degeneracy of different bounces, discussed in section~\ref{sect:exD}
is now lost (compare with Figure~\ref{fig:ExDS}). The reason for this is discussed below as it is better understood in the tunneling potential approach, to which we turn next.

\begin{figure}[t!]
\begin{center}
\includegraphics[width=0.5\textwidth]{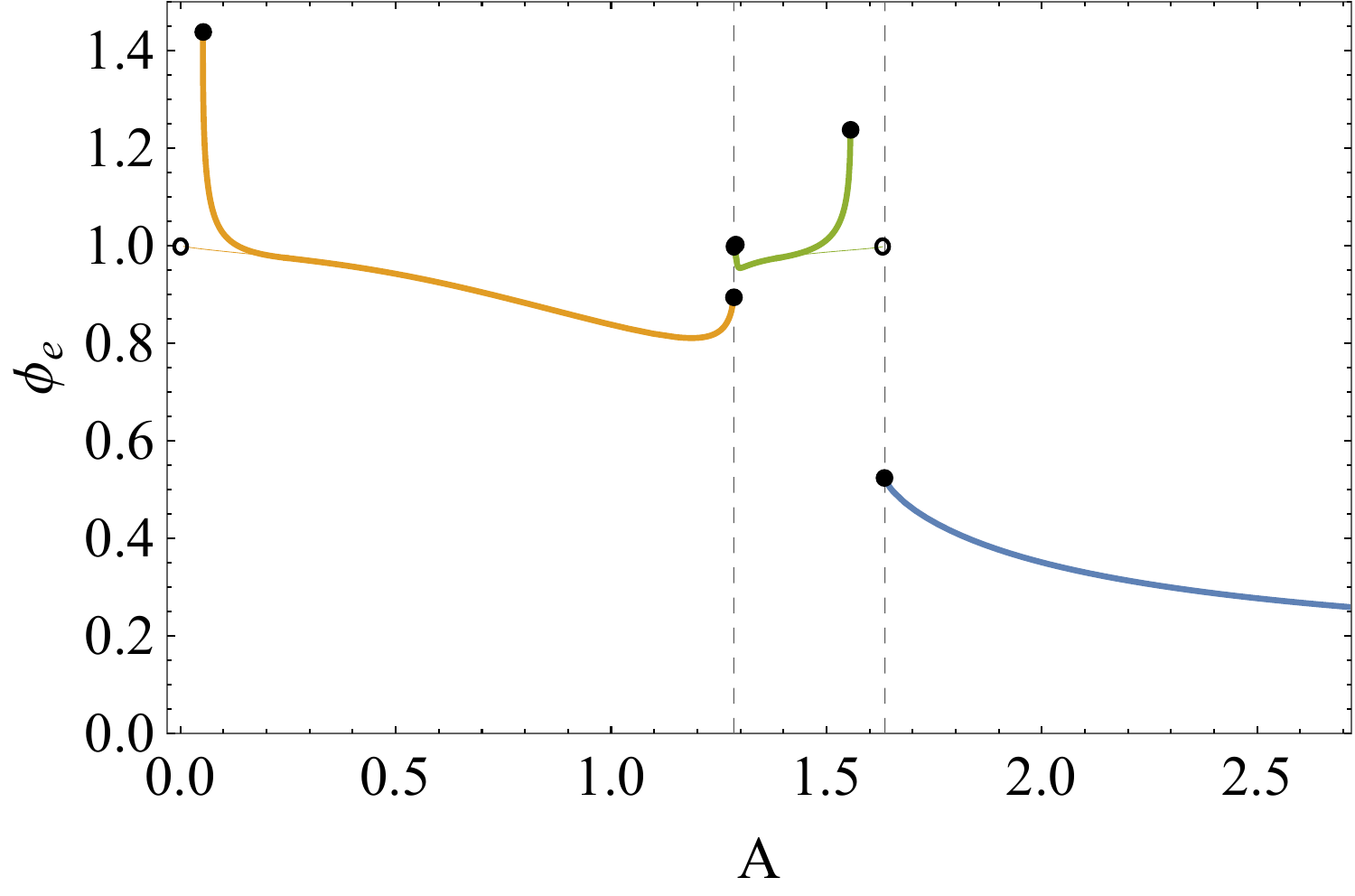}
\end{center}
\caption{Same as Figure~\ref{fig:ExDVt} but for the regularized potential of this section.
Thin lines correspond to the original potential $V$, as in Figure~\ref{fig:ExDVt}. Empty-circle points indicate bounces for $V$ and black points for $V_R$.
\label{fig:ExDVtReg}
}
\end{figure}

\begin{figure}[t!]
\begin{center}
\includegraphics[width=0.5\textwidth]{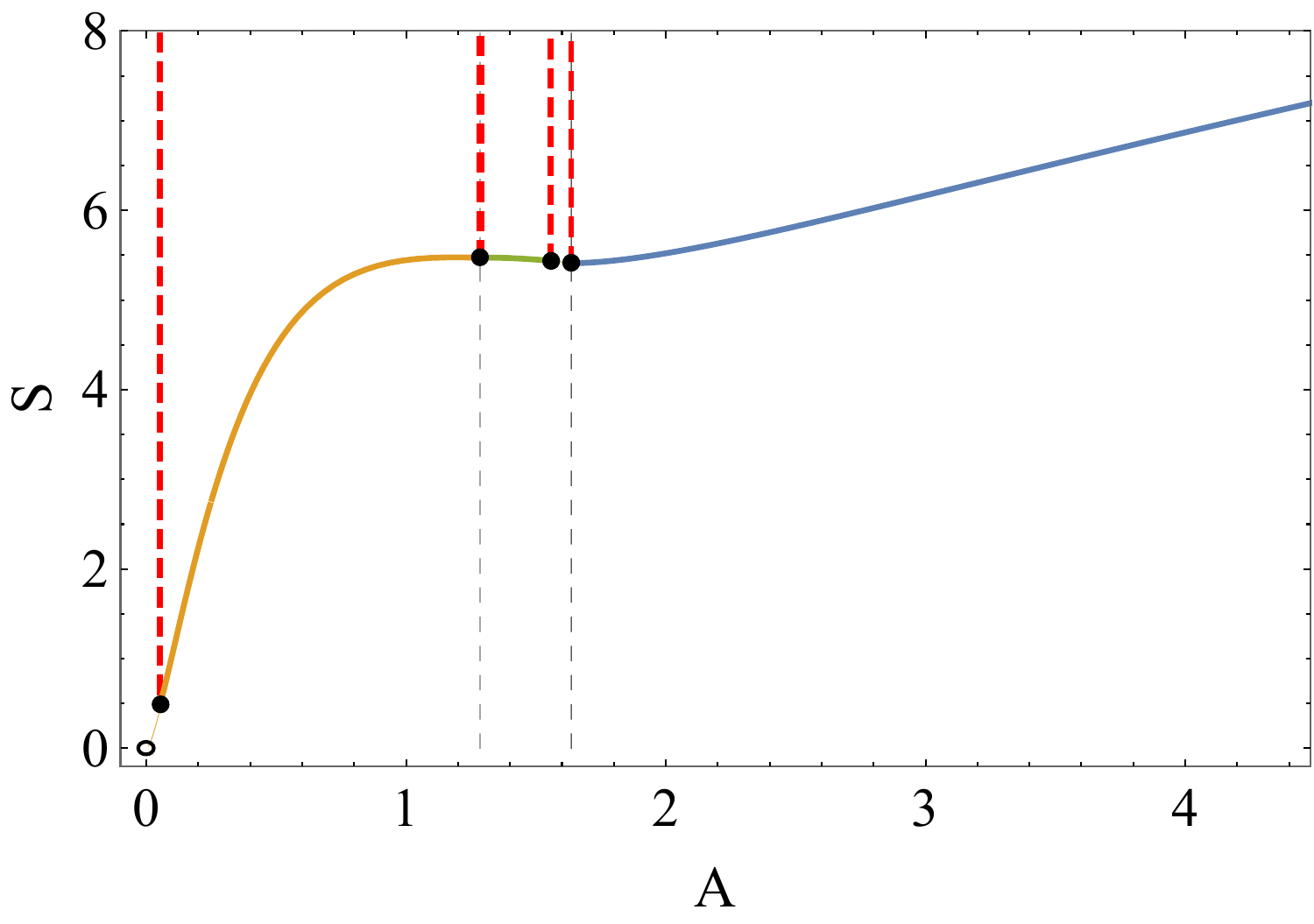}\\
\includegraphics[width=0.46\textwidth]{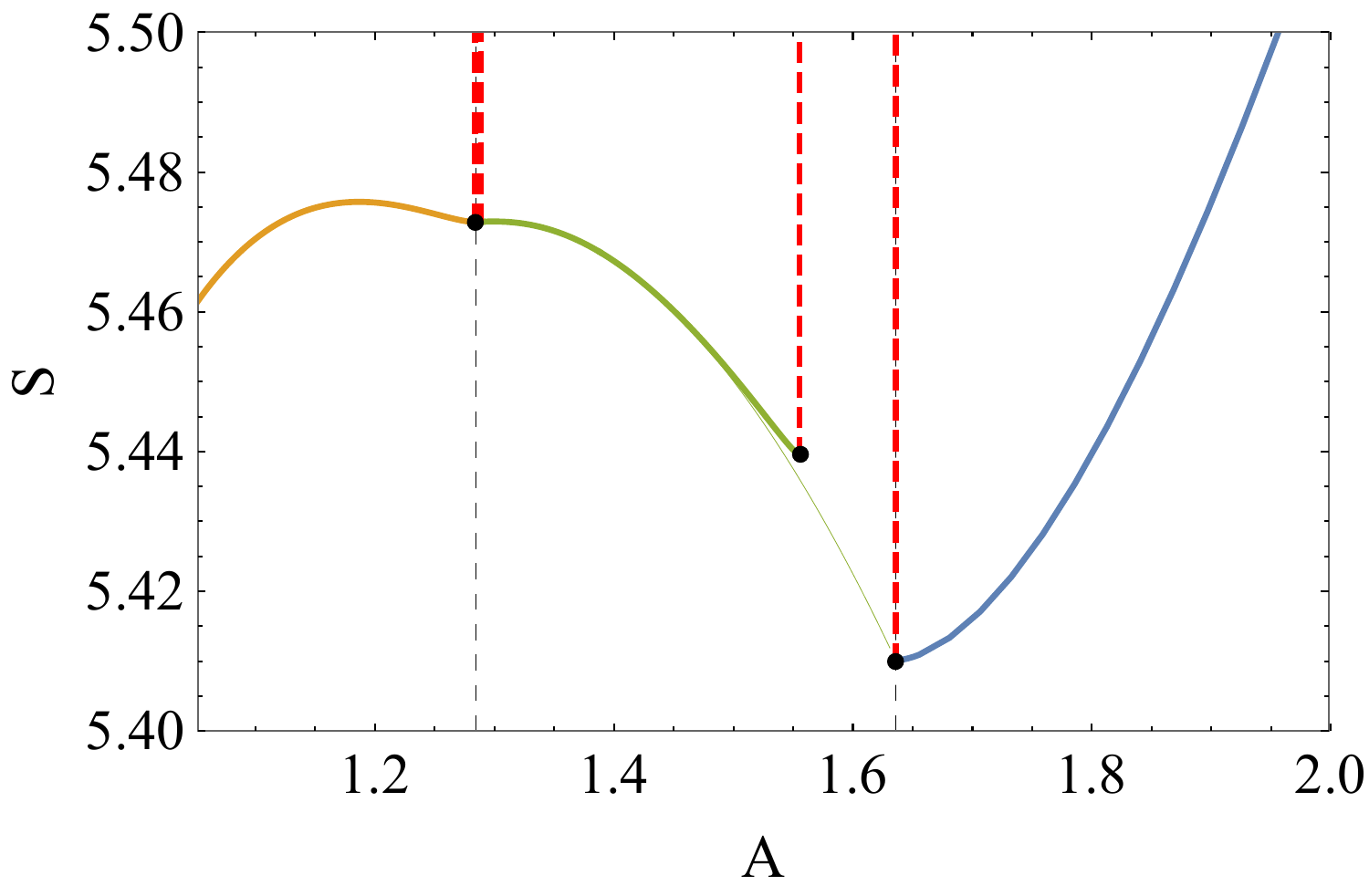}
\includegraphics[width=0.49\textwidth]{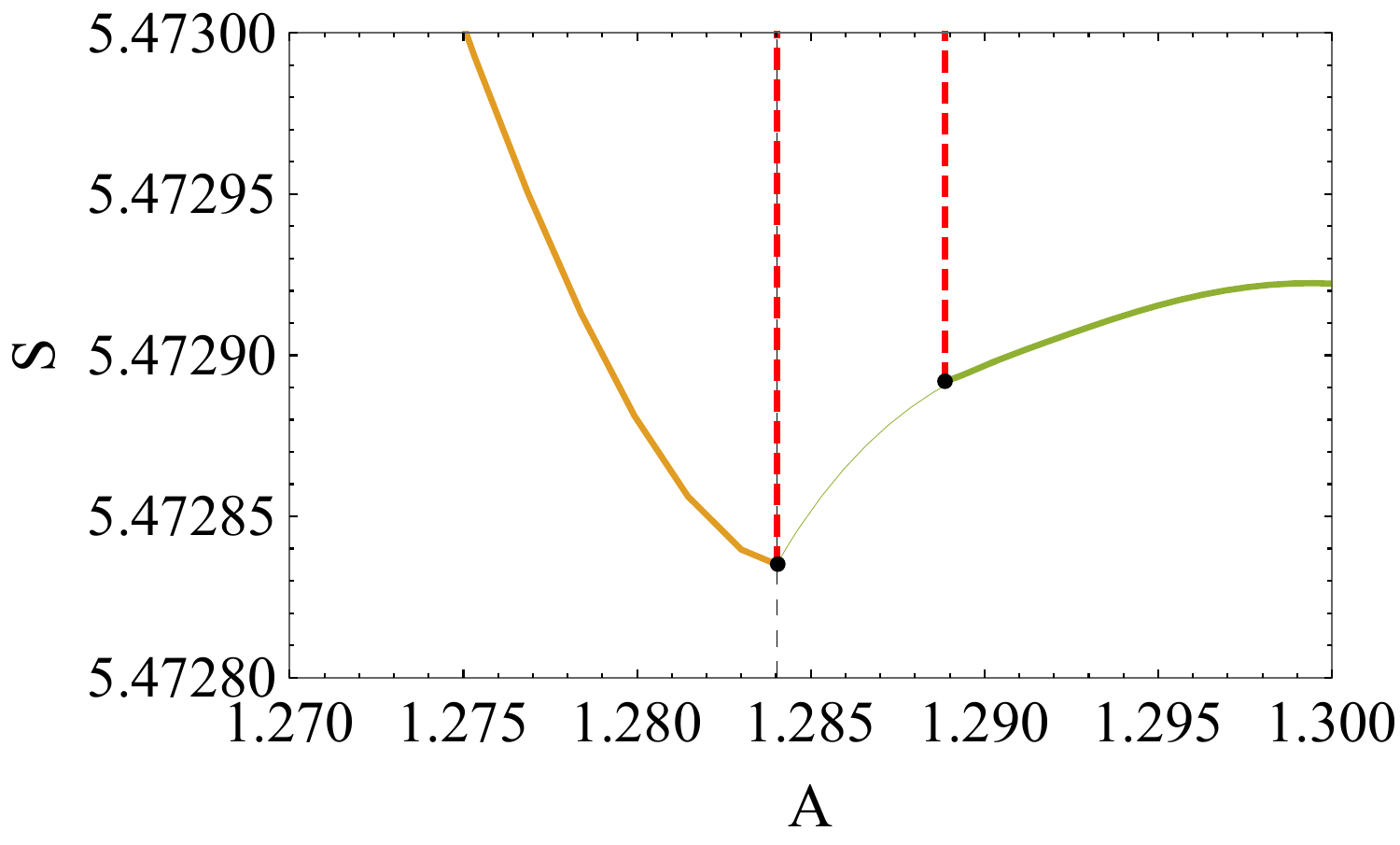}
\end{center}
\caption{Tunneling action for $V_t(A;\phi)$ solutions for the regularized potential considered in this section, using the same color coding of previous figures. 
Lower plots are zoom-ins of the top plot.
Dots mark the location of bounces and correspond to local extremals
of $S(A)$. Red dashed lines bracket field intervals where $S(A)\to\infty$.
\label{fig:ExDVtSreg}
}
\end{figure}

If we use the same $A$ parametrization of $V_t$ solutions used for section~\ref{sect:exD},
we find that the function $\phi_e(A)$ is modified by the regularization of the potential as shown in Figure~\ref{fig:ExDVtReg} while the tunneling action $S(A)$ is given in
Figure~\ref{fig:ExDVtSreg}, with the top plot giving the overall picture and the the two lower plots zooming in on particular regions. In these plots we follow the same color coding and line-type prescriptions of previous figures. 
Bounces for $V$ are marked by empty circles and for $V_R$ with black dots. For $\phi_e\leq \phi_x$ nothing changes, as expected. For three
small intervals of $A$ [$A\in (0,0.052), A\in(1.284,1.289), A\in(1.556, 1.636)$] $\phi_e\to\infty$ with the $V_t(A;\phi)$ solutions diverging to $-\infty$. Such divergent solutions
are similar to bubble-of-nothing (BoN) solutions found in extra-dimensional models  when a 4-dimensional effective description is used (with $\phi$ being a modulus field) \cite{BoN0,BoN1,BoN2}. However, in this case these solutions have infinite action (and we have indicated this in Figure~\ref{fig:ExDVtSreg}  using vertical red-dashed lines)\footnote{This breaks the continuity of $S(A)$ in some intervals. In the case of bubble-of-nothing solutions, gravity renders finite the action across such intervals and one gets a continuous action again \cite{BoN2}.}. The reason for this can be understood as follows: for such solutions,
$V$ is negligible compared to $V_R$ for large enough $\phi$ and $V_t\simeq -C\phi^3$. This leads to a constant action density that integrates to $\infty$. For the potential
of section~\ref{sect:exD} such cubic solutions were cut off at $\phi_e=1$ leading to a finite
action that decreased towards $S=0$.  Thus, the regularization of the potential achieves a finite decay rate (for the lowest point of the curve in Figure~\ref{fig:ExDVtSreg}), as one would  expect.

\section{Some General Lessons\label{sect:genless}}

\subsection{2n+1 bounces}
By examining in this work pseudo-bounce false-vacuum decays in several scalar potentials already discussed in the literature, 
we have uncovered a structure of decays more general than the one usually 
considered. Instead of a single bounce solution that could be found by standard overshoot-undershoot algorithms, we find that the more general case consists
of $2n+1$ bounce solutions, with bounces alternating with antibounces. As explained before, bounces have the standard behavior with respect to overshoot/undershoot searches: if the initial starting field value of the bounce, $\phi_e$, is increased (decreased) one gets an overshot (undershot). Antibounces have the opposite behavior.

These $2n+1$ solutions organize themselves in $n$ pairs of bounce-antibounce solutions connected by pseudo-bounce configurations, and a single bounce connected by pseudo-bounces to a configuration of infinite radius and infinite action. Examples of 
such structure in the plane $(\phi_e,r_i)$ are shown in Figure~\ref{fig:ExDri} (with 
$n=2$), Figure~\ref{fig:Sasakiphieri} (with $n=\infty$), and Figure~\ref{fig:ExDri_reg} (with $n=2$). Even if most potentials with a false vacuum have just one bounce,
it is important to keep in mind that more general cases are possible. Existing numerical codes \cite{Wainwright,MOS,Hollik,Athron,Sato,Guada,Bardsley,Maggie,Hua} are designed to find just one bounce in single-field cases and would miss the rich decay structure uncovered in this paper if confronted with potentials like the ones we have considered.

Our discussion applies to single-field scalar potentials.  If $V(-\phi)=V(\phi)$, there are two (degenerate) true minima and multi-pass pseudo-bounce solutions appear, but these
typically cost more action and can be neglected. If the potential is not symmetric but it features two minima, to the left and right of the false vacuum $\phi_+=0$, decays towards $\phi>\phi_+$ or towards $\phi<\phi_+$ can be considered separately, each with its own potential. Finally, exploring the decay structure of multi-field potentials would require a separate study beyond this work.

\begin{figure}[t!]
\begin{center}
\includegraphics[width=0.6\textwidth]{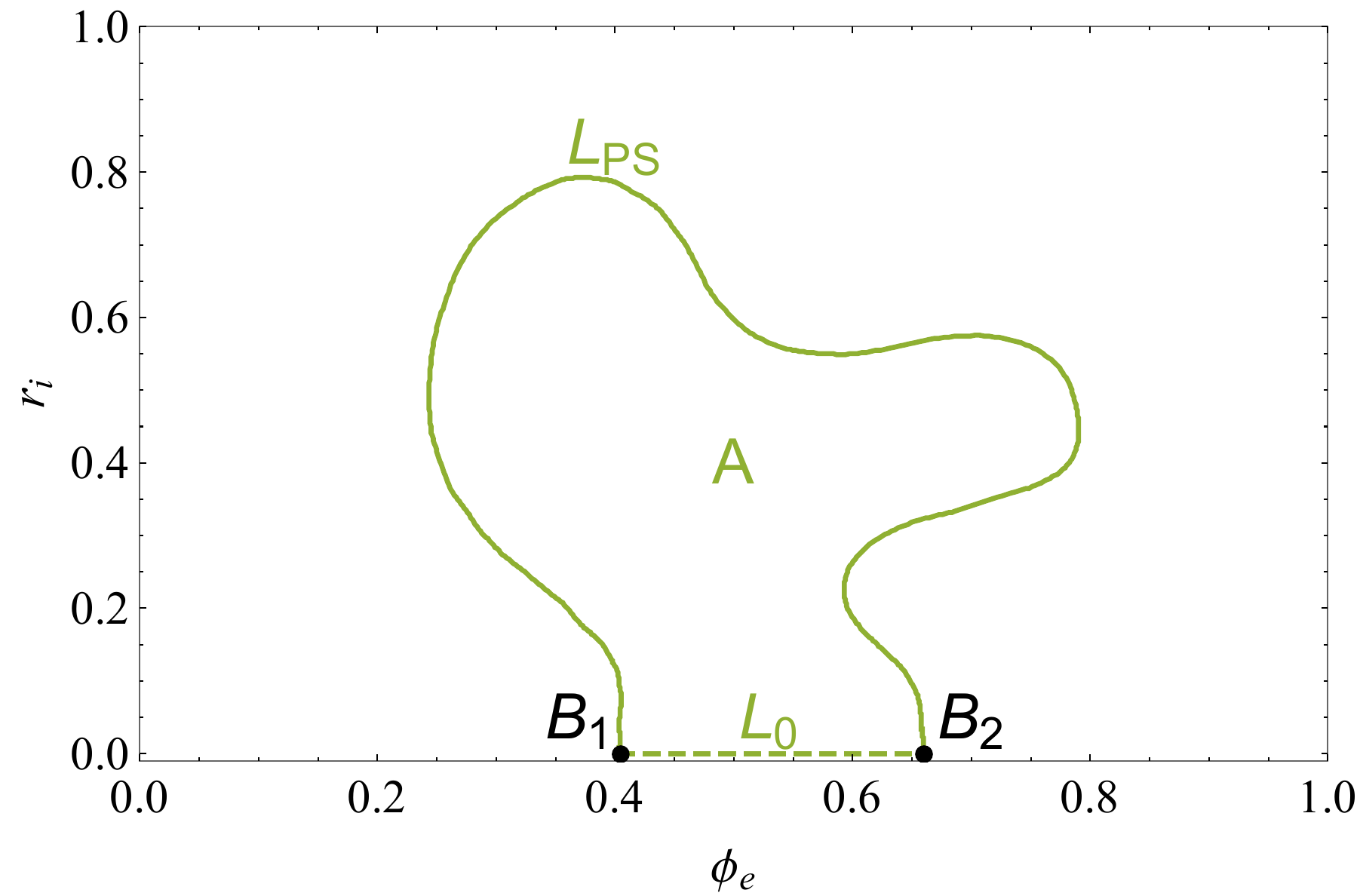}
\end{center}
\caption{Schematic representation of a bounce-antibounce pair ($B_1$,$B_2$) joined by a line $L_{PS}$ of pseudo-bounce solutions. The false vacuum is assumed to sit at $\phi=0$.
\label{fig:stokes}
}
\end{figure}

\subsection{The relative actions of bounce-antibounce pairs}
For a given bounce-antibounce pair (joined by a pseudo-bounce line), some general statements can be made about the relative value of their respective actions. For this we use the relation for pseudo-bounce actions (\ref{dSdphie}):
\be
\frac{dS}{d\phi_e}=\frac{\pi^2}{2}r_i^4(\phi_e) V'(\phi_e)\ ,
\label{dSdphie2}
\ee
that we used already in section~\ref{sect:exD}. Consider a bounce-antibounce pair connected by a pseudo-bounce line, as in the examples considered in this paper and shown schematically in Figure~\ref{fig:stokes}. If the $A$ region corresponds to undershots (overshots), $B_1$ ($B_2$) is the antibounce. Irrespective of the shape of the pseudo-bounce line ($L_{PS}$ in the figure), one can prove that the action of the bounce or antibounce with larger $\phi_e$ ($B_2$ in the figure) has lower action. Using (\ref{dSdphie2}) along the pseudo-bounce line, we get
\be
\Delta S_{21}\equiv S_{B_2}-S_{B_1} = \frac{\pi^2}{2}\int_{L_{PS}}r_i^4(\phi_e)V'(\phi_e)d\phi_e\ ,
\ee
with the integral taken along $L_{PS}$ from $B_1$ to $B_2$. We can prolong $L_{PS}$ along $L_0$ into the closed circuit $\partial A$ (the boundary of the region $A$) and use Stoke's theorem\footnote{For $(x,y)=(\phi_e,r_i)$ with the vector field $\vec{E}\equiv(-\pi^2r_i^4V'(\phi_e)/2,0)$, so that $\nabla\times \vec E=2\pi^2r_i^3V'(\phi_e)$.} and the fact that $r_i=0$ at $L_0$ to obtain
\be
\Delta S_{21} = 
-\frac{\pi^2}{2}\oint_{\partial A}r_i^4(\phi_e)V'(\phi_e)d\phi_e
= 2\pi^2\int_{A}r_i^3V'(\phi_e)dr_id\phi_e <0\ ,
\ee
where the overall sign change is due to the clockwise sense of tracing $\partial A$ and
the last inequality follows from $V'(\phi_e)<0$. If we had another bounce-antibounce pair $(B_1',B_2')$ inside the interval $(B_1,B_2)$ connected by a pseudo-bounce stretching inside $A$,
the same logic would give us the inequalities $\Delta S'_{21}\equiv S_{B_2'}-S_{B_1'}<0$ and $|\Delta S_{21}|>|\Delta S'_{21}|$. The examples of previous sections confirm this behavior.
On the other hand, comparing the action of bounces not connected by pseudo-bounce lines requires going beyond pseudo-bounce configurations. We face the same complication in the next subsection.

\subsection{A slice of the Euclidean landscape}

\begin{figure}[t!]
\begin{center}
\includegraphics[width=\textwidth]{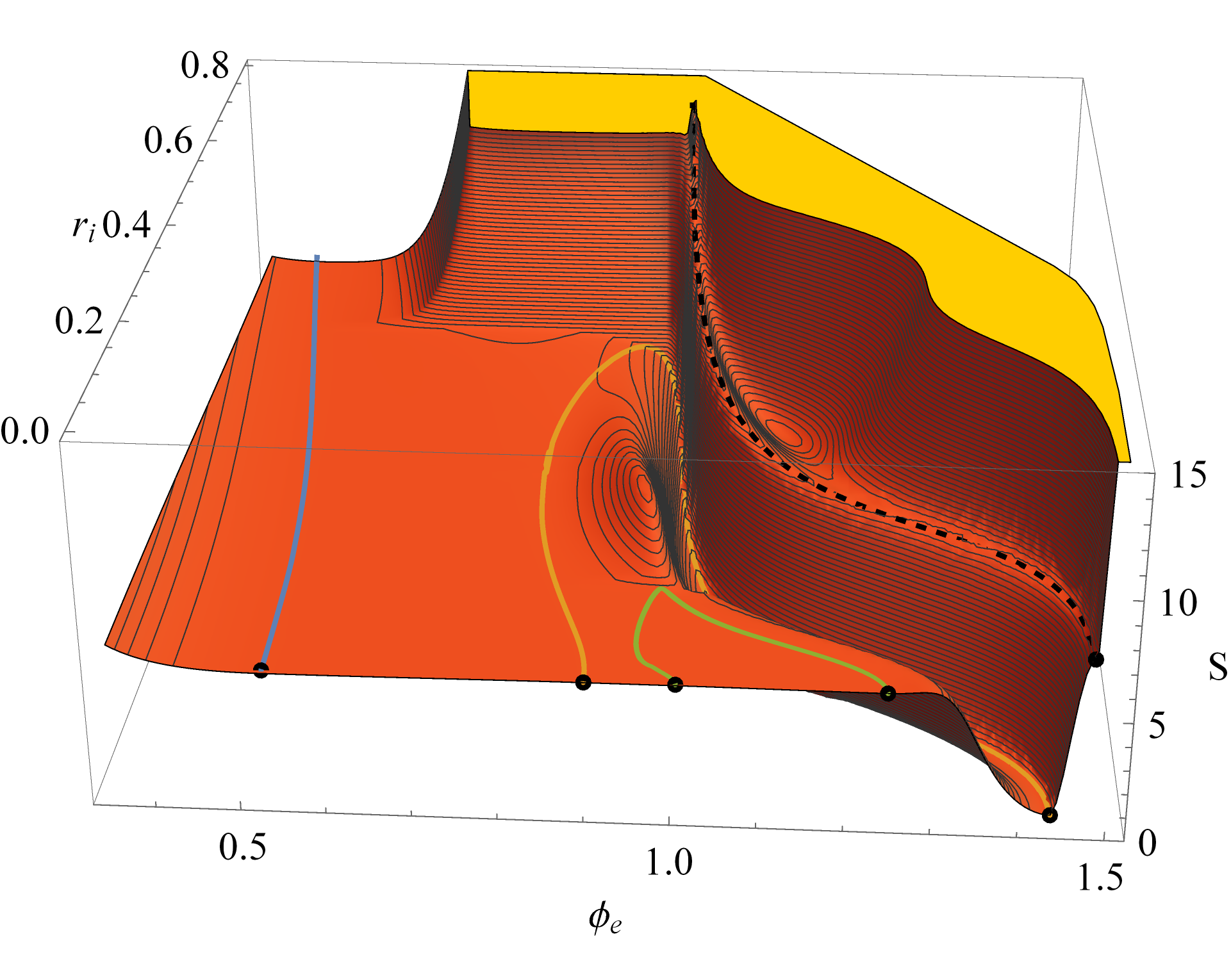}
\end{center}
\caption{
\label{fig:3dLandscape} Action landscape with pseudo-bounce valleys (colored lines) for the model of section~\ref{sect:exDreg}. 
}
\end{figure}  

\begin{figure}[t!]
\begin{center}
\includegraphics[width=\textwidth]{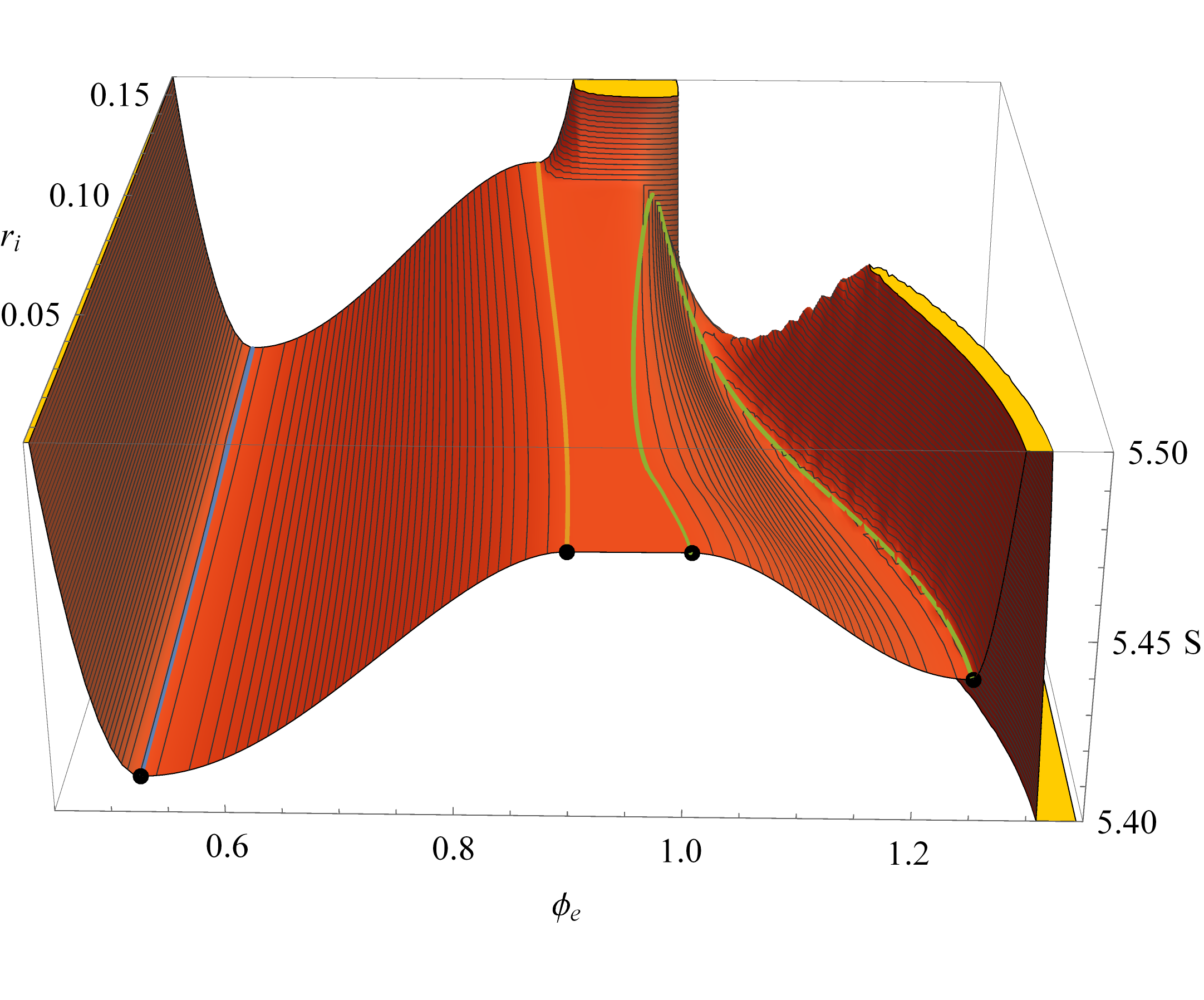}
\end{center}
\caption{
\label{fig:3dLandscapeZoom} Zoomed-in region of Figure~\ref{fig:3dLandscape}. 
}
\end{figure} 

In this subsection we provide a glimpse of how pseudo-bounces appear along valleys in a landscape that gives the action for different field configurations. The space of field configurations is infinite dimensional, of course, but we just want a two-dimensional slice that passes through our pseudo-bounce solutions. 

Consider as an example our regularised model of section~\ref{sect:exDreg}. Figure~\ref{fig:ExDri_reg} could be such a slice, with field configurations labelled by $\phi_e$ and $r_i$. An immediate difficulty is that generic points in that plane correspond to overshots or undershots, and such configurations are not acceptable solutions describing vacuum decay. We could try to truncate them following some prescription, ensuring also that the corresponding critical bubbles (their associated 3-dimensional configurations, given by the slice of the 4-dimensional configuration at zero Euclidean time) have zero energy. A simpler alternative is to associate
to each point $(\phi_e,r_i)$ a field configuration that is an interpolation
of two nearby pseudo-bounce configurations, chosen as follows. For a generic $(\phi_e,r_i)$ point
find the two pseudo-bounces, $\phi_{1,2}(r)$, with the same $r_i$ that
satisfy $\phi_1(r_i)\leq \phi_e\leq \phi_2(r_i)$. If the point $(\phi_e,r_i)$ lies to the left of the first pseudo-bounce line, then 
$\phi_1$ is taken to be 0.
We then define the field configuration
\be
\phi(r)\equiv\alpha \phi_1(r) + (1-\alpha)\phi_2(r)\ .
\ee
It also has a core of radius $r_i$, with
$\phi_e=\phi(r_i)$ interpolating between the two pseudo-bounce lines if $\alpha$ is varied with $0\leq \alpha\leq 1$. To ensure
that the energy associated to this configuration is zero, we simply rescale
it as
\be
\phi_a(r)\equiv \phi(a r)\ ,
\ee
choosing $a$ so as to get zero energy.\footnote{A simple scaling argument gives $a^2=-(\int dr\ r^2 V )/(\int dr\ r^2 \dot\phi^2/2)$.} Although the rescaling also changes the size of the inner core to $r_i/a$, we are free to keep using the initial
$r_i$ parameter to label the configuration. 

If we then calculate the Euclidean action for such configurations in the $(\phi_e,r_i)$ plane, we obtain Figures~\ref{fig:3dLandscape} and ~\ref{fig:3dLandscapeZoom} (a zoomed-in region of the former). We have marked the pseudo-bounce lines using the same color coding used in section~\ref{sect:exDreg}. The black dots mark the true bounces (and antibounces). We see that pseudo-bounce lines indeed follow valleys
in this landscape. We also see the 1-pass pseudo-bounce (dashed black line in Figure~\ref{fig:3dLandscape}), which follows a ridge instead. This agrees with the common expectation that these solutions have more than one negative mode. 
Multi-pass pseudo-bounces with higher number of passes also follow ridges and have even higher actions.

\subsection{IR vs. UV labels for pseudo-bounce families}

In the main text we have used two different kinds of labels for the families of pseudo-bounces we have studied. One, $r_i$ (and/or $\phi_e$), is natural to use in the Euclidean approach
(as they are the size and field value of the core  of the pseudo-bounce field configuration); and the other, $A$, is natural to use in the tunneling potential approach (where it appears
as a free parameter in the low field expansion of $V_t$). 

A more important difference between the two labels is the following. The field value $\phi_e$ is the largest value taken by the field
configuration of the pseudo-bounce, and it is realized at the smallest radial distances, up to $r_i$. Thus, $\phi_e$ and $r_i$ can be thought of as UV labels. 
Indeed, they are sensitive to large field values and short distances and would be modified by the possible presence of new physics at such scales. 

On the other hand, as $A$ arises from the field expansion of the pseudo-bounce solution near
the false vacuum, it can be thought of as an IR label, sensitive to low-field
excursions from the false vacuum which correspond to large distances from the center of the field configuration.
It is intriguing that the behavior of the tunneling action is smoother and simpler if one uses
the IR label instead of the UV ones, as we have seen in all the examples we have examined.
From this point of view, the UV label seems to be the natural one to describe these decay configurations.

%%%%%%%%%%%%%%%%%%%%%%%%%%%%%%%%%%%%%%%%
\section*{Acknowledgments\label{sec:ack}} 
%%%%%%%%%%%%%%%%%%%%%%%%%%%%%%%%%%%%%%%%
We thank Maggie M\"uhlleitner and her collaborators in \cite{Maggie} for analyzing the example of \cite{EK} we discuss in section~\ref{sect:exD}, finding a bounce
different from the analytic one we expected, and bringing it to our attention.
The work of J.R.E. has been supported by the IFT Centro de Excelencia Severo Ochoa
CEX2020-001007-S and by PID2022-142545NB-C22 funded by MCIN/AEI/10.13039/ 501100011033
and by “ERDF A way of making Europe”. 
TK acknowledges support by the Deutsche Forschungsgemeinschaft (DFG, German Research Foun-
dation) under Germany’s Excellence Strategy – EXC 2121 “Quantum Universe” - 390833306. 

\appendix
\section{Singular bounce of section~\ref{Sasaki}}
We present here the singular bounce solution for the potential discussed in Section~\ref{Sasaki}, with $\alpha=1,\phi_\star=1, m_1=m_2=1$. The field profile in the different regions of the potential can be obtained analytically \cite{Sasaki}. For the exponential region of the potential (with $\phi\geq 0$ and $0\leq r\leq r_0$)
\be
\phi_B(r)=-\log r\ .
\label{phiexp}
\ee
For $\phi_2\leq\phi\leq 0$ (and $r_0\leq r\leq r_1$),
\be
\phi_B(r)=-2+\frac{c_J}{r}J_1(r)+\frac{c_Y}{r}Y_1(r)\ ,
\ee
where $J_1$ and $Y_1$ are the Bessel functions of the first and second kind, respectively. For $\phi_M\leq\phi\leq\phi_2$, (and $r>r_1$)
\be
\phi_B(r)=\phi_M+\frac{c_I}{r}I_1(r)+\frac{c_K}{r}K_1(r)\ ,
\ee
where $I_1$ and $K_1$ are the modified Bessel functions of the first and second kind, respectively.

\begin{figure}[t!]
\begin{center}
\includegraphics[width=0.45\textwidth]{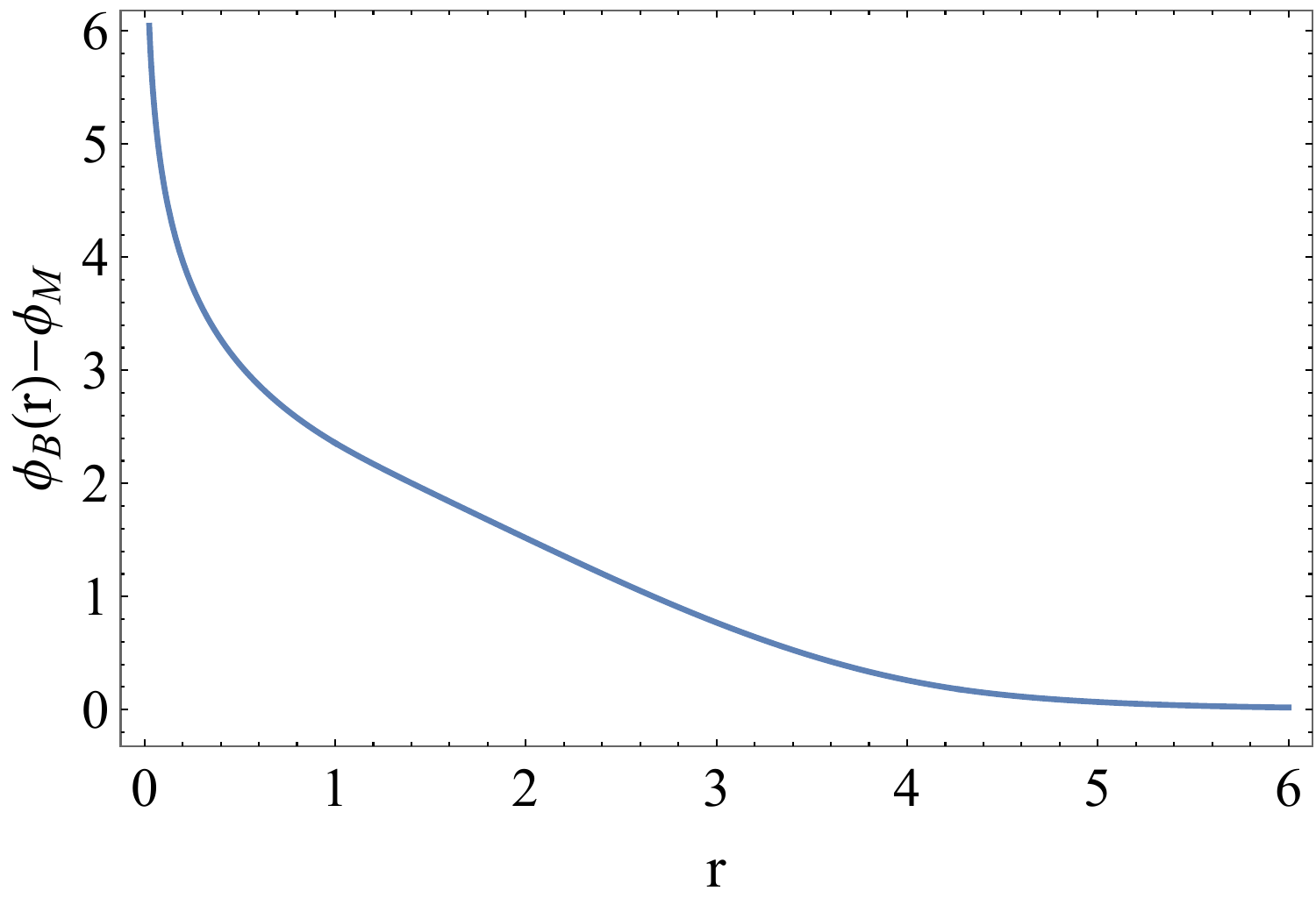}
\end{center}
\caption{
\label{fig:SasakiBounce} Field profile of the singular bounce of Section~\ref{Sasaki}, with $\phi_B(0)=\infty$.
}
\end{figure}  

We match $\phi_B$ and $\phi_B'$ at $r_0$ and $r_1$ and impose
the boundary conditions $\phi_B(0)=\infty$ and $\phi_B(\infty)=\phi_M$.
The four integration constants $c_J, c_Y, c_I, c_K$, plus $\phi_M$ and $r_{0,1}$ are then determined by the relations
\be
c_J=\frac{\pi}{2}\left[2Y_0(1)-3Y_1(1)\right]\ ,
\quad
c_Y=\frac{2-c_J J_1(1)}{Y_1(1)}\ , \quad c_I=0\ ,
\ee
and
\be
c_K=\frac{4Y_1(r_1)-2r_1(\phi_M+2)Y_1(1)+2c_J[J_1(r_1)Y_1(1)-J_1(1)Y_1(r_1)]}{2K_1(r_1)Y_1(1)}\ 
\ee 
plus
\be
\phi_M=2\left[\frac{c_J J_1(r_1)}{r_1}+\frac{c_Y Y_1(r_1)}{r_1}-1\right]\ ,\quad r_0=1  \ .
\ee
Finally, $r_1$ is found numerically by solving 
the relation
\be
\frac{r_1 \left[J_0(r_1)K_1(r_1)-J_1(r_1)K_0(r_1)\right]-4J_1(r_1)K_1(r_1)}{r_1 \left[Y_0(r_1)K_1(r_1)-Y_1(r_1)K_0(r_1)\right]-4Y_1(r_1)K_1(r_1)}+\frac{c_Y}{c_J}=0\ .
\ee
With our choice of parameters one gets $\phi_M=-2.358$ and $r_1=4.275$. The field profile of the singular bounce is shown in Figure~\ref{fig:SasakiBounce}. The Euclidean action for this bounce is $S_0=130.44$.

\section{\bma{V_t} Crossing \label{app:Vtcross}}
The low-field expansions of $V_t(A;\phi)$, see eqs.~(\ref{VtA}) and (\ref{VtW}),
imply that, for low enough $\phi$, $V_t(A_1;\phi)>V_t(A_2;\phi)$ 
if $A_1>A_2$. If the $V_t$ solutions do not cross at higher values of $\phi$, then $\phi_e(A)$ would be a monotonically decreasing function of $A$, as a higher $V_t$ hits $V$ earlier. Whenever $\phi_e(A)$ is non monotonic (as in our examples, see Figures~\ref{fig:ExDVt} and \ref{fig:ExDVtReg}) $V_t$ solutions cross.\footnote{The converse is not true: solutions might cross twice and give a monotonic $\phi_e(A)$.}

When any two solutions $V_t(A_1;\phi)$ and $V_t(A_2;\phi)$ cross
(say at $\phi_x$) we can construct a ``mixed solution'' that lowers the action of one of the two. Simply, from $\phi_x$ onwards, follow the solution that gives the lowest contribution to the action in the field interval beyond the crossing point. So, writing (for $a=1,2$)
\be
S(A_a)=\int_{0}^{\phi_{ea}}s_a(\phi)d\phi=\int_{0}^{\phi_x}s_a(\phi)d\phi+\int_{\phi_x}^{\phi_{ea}}s_a(\phi)d\phi\ ,
\ee
if
\be
\int_{\phi_x}^{\phi_{e2}}s_2(\phi)d\phi > \int_{\phi_x}^{\phi_{e1}}s_1(\phi)d\phi\ ,
\ee
then the mixed solution 
\be
V_{t21}(\phi)=\left\{
\begin{matrix}
V_t(A_2;\phi)&,& \mathrm{for}\; \phi\le \phi_x\\
V_t(A_1;\phi)&,& \mathrm{for}\; \phi\ge \phi_x
\end{matrix}
\right.
\ee
lowers the action $S(A_2)$. We have checked, in the example of section~\ref{sect:exD}, that the effect of this on the function $S(A)$ is to make it monotonic, removing the maximum and kinks of the action $S(A)$ shown in Figure~\ref{fig:ExDVtS}. Although this is an interesting fact, the mixed solutions have a problematic physical interpretation, as discussed below.

The mixed solutions are not true solutions (they do not satisfy the EoM for $V_t$  at $\phi_x$), but this by itself is not a problem: pseudo-bounces are not solutions either. The Euclidean profile of a mixed solution follows simply from the dictionary between Euclidean and $V_t$ formalisms given in section~\ref{sec:Vt}. In particular, using (\ref{rVt}), the different values of
$V_t'(\phi_x)$ for the two $V_t$ solutions crossing at $\phi_x$ lead to different values of $r$ in the unmixed Euclidean profiles [say $r_a$ for $V_t(A_a;\phi)$]. Between these two $r$ values, the field stays constant and equal to $\phi_x$. Continuing with the previous example with $A_1$ and $A_2$ solutions crossing, the full profile
of the mixed solution will consist of the pseudo-bounce profile of solution $A_1$ from $r=0$ to $r_1$, a constant $\phi_x$ from $r_1$ to $r_2$ and finally the pseudo-bounce profile of solution $A_2$, from $r_2$ to $\infty$. Notice that such profile only makes sense if $r_2>r_1$
(and thus $|V_{t1}'|>|V_{t2}'|$, which is the case in our example). 

Such mixed solution lives in 4d Euclidean space and its slice at zero Euclidean time gives the 
3d profile of the nucleated bubble. The trouble is that the energy
of such bubble is not zero and therefore, this profile is not really describing a vacuum decay process. The proof is as follows. The total energy integral is
\be
E=4\pi\int_0^\infty dr\ r^2\left[\frac12\dot\phi^2+V(\phi)\right]\equiv \int_0^\infty dr\ e(r)\ .
\ee
For the mixed solution of our example  this integral splits as\footnote{The kinks in the profile of the mixed solution do not contribute delta-function terms to the energy density.}
\be
E=\frac{4\pi}{3}r_{i1}^3V(\phi_{e1})+\int_{r_{i1}}^{r_1} dr\ e_1(r)
+ \frac{4\pi}{3}(r_2^3-r_1^3)V(\phi_x)+\int_{r_2}^\infty dr\ e_2(r)\ ,
\label{Esplit}
\ee
where the first term corresponds to the inner core of the pseudo-bounce solution $1$, with radius $r_{i1}$; the second to the profile of the solution $1$; the third to the spherical shell with constant $\phi_x$ between $r_1$ and $r_2$; and the fourth to the profile of the solution $2$. Next we use the fact that $e(r)$ is a total derivative on solutions of the EoM [thus in the intervals $(r_{i1},r_1)$ and $(r_2,\infty)$]\footnote{Notice that (\ref{etotalder}) can be written as $e(r)=d[{\mathrm{Vol}(S_{3,r})}V_t]/dr$, giving another interpretation to the tunneling potential: as a function of $r$, $V_t(r)$ is the mean energy density inside the critical bubble up to radius $r$.}
\be
e(r)=\frac{d}{dr}\left[\frac{4\pi r^3}{3}\left(V-\frac12\dot\phi^2\right)\right]\ ,
\label{etotalder}
\ee
to calculate explicitly the integrals in (\ref{Esplit}) and 
we arrive at 
\be
E=\frac{2\pi}{3}\left(r_{i1}^3\dot\phi_{e1}^2-r_1^3\dot\phi_1^2+r_2^3\dot\phi_2^2\right)\ .
\ee
For the pseudo-bounce inner core we have  $\dot\phi_{e1}=0$ and the first term vanishes. However, for the crossing point, from $V_t(A_1;\phi_x)=V_t(A_2;\phi_x)$
and $V_t=V-\dot\phi^2/2$ we have $\dot\phi_1=\dot\phi_2$, which looks promising, but $r_1\neq r_2$ and thus $E\neq 0$. Therefore, only if the crossing of solutions occurs with the same $V'_t(\phi_x)$  would the energy of the bubble be zero. However, solutions with the same $V$, $V_t$ and $V_t'$ at $\phi_x$ also have the same $V_t''$ [fixed by the EoM (\ref{EoMVt})] and are thus the same single solution. We are thus forced to discard mixed solutions.  We have not found an alternative physical interpretation that is useful.

\end{document}